\documentclass[10pt]{article}
\usepackage{amssymb,amsbsy}
\usepackage{amsmath}
\usepackage{amsthm}
\usepackage{graphics,graphicx}
\usepackage[T1]{fontenc}
\usepackage{color}

\newcommand{\beq}{\begin{equation}}
\newcommand{\eeq}{\end{equation}}
\newcommand{\lb}{\label}
\newcommand{\beqar}{\begin{eqnarray}}
\newcommand{\eeqar}{\end{eqnarray}}
\newcommand{\bit}{\begin{itemize}}
\newcommand{\eit}{\end{itemize}}
\newcommand{\benum}{\begin{enumerate}}
\newcommand{\eenum}{\end{enumerate}}
\newcommand{\barr}{\begin{array}}
\newcommand{\earr}{\end{array}}
\def\ds{\displaystyle}
\newcommand\eq[1]{(\ref{#1})}
\newtheorem{theorem}{Theorem}[section]

\newtheorem{lemma}{Lemma}[section]

\def\scalp{\mbox{\boldmath $\, \cdot \,$}}

\newcommand{\deriv}[2]{\frac{\partial #1}{\partial #2}}

\def\XXint#1#2#3{{\setbox0=\hbox{$#1{#2#3}{\int}$}
   \vcenter{\hbox{$#2#3$}}\kern-.5\wd0}}

\def\b0{\mbox{\boldmath $0$}}
\def\ba{\mbox{\boldmath $a$}}
\def\bb{\mbox{\boldmath $b$}}

\def\bm{\mbox{\boldmath $m$}}

\def\bq{\mbox{\boldmath $q$}}

\def\bA{\mbox{\boldmath $A$}}
\def\bB{\mbox{\boldmath $B$}}

\def\bI{\mbox{\boldmath $I$}}

\def\bQ{\mbox{\boldmath $Q$}}

\def\bS{\mbox{\boldmath $S$}}

\def\bX{\mbox{\boldmath $X$}}

\newcommand{\bsigma}{\mbox{\boldmath $\sigma$}}

\def\f0{\ensuremath{\mathbb{O}}}

\newcommand{\tr}{\mathop{\mathrm{tr}}}

\newcommand{\sign}{\mathop{\mathrm{sign}}}

\def\Sym{\ensuremath{\mathsf{Sym}}}
\newcommand{\Reals}{\ensuremath{\mathbb{R}}}

\def\AMEC{{\it Acta Mech.}\ }

\def\CMAME{{\it Comput.\ Method.\ Appl.\ M.}\ }

\def\IJMS{{\it Int.\ J.\ Mech.\ Sci.}\ }

\def\IJP{{\it Int.\ J.\ Plasticity}\ }
\def\IJSS{{\it Int.\ J.\ Solids Struct.}\ }

\def\JAM{{\it ASME J.\ Appl.\ Mech.}\ }

\def\JMPS{{\it J.\ Mech.\ Phys.\ Solids}\ }

\def\MMS{{\it Math.\ Mech.\ Solids}\ }
\def\MOM{{\it Mech.\ Materials}\ }

\def\PRSL{{\it Proc.\ R.\ Soc.\ Lond.}\ }

\title{Yield criteria for quasibrittle and frictional materials: \\ a generalization to surfaces with corners}

\author{Andrea Piccolroaz and Davide Bigoni
\\
\\
{\it Dipartimento di Ingegneria Meccanica e
Strutturale, Universit\`a di Trento,}
\\ 
{\it Via Mesiano 77, I-38050 Trento, Italia}
}

\begin{document}

\maketitle

\begin{abstract}
\noindent
Convexity of a yield function (or phase-transformation function) and its relations to convexity of the corresponding yield surface 
(or phase-transformation surface) is essential to the invention, definition and comparison with experiments of new yield (or phase-transformation) 
criteria.
This issue was previously addressed only under the hypothesis of smoothness of the surface, but yield surfaces with corners (for instance, the 
Hill, Tresca or Coulomb-Mohr yield criteria) are known to be of fundamental importance in plasticity theory.
The generalization of a proposition relating convexity of the function and the corresponding surface 
to nonsmooth yield and phase-transformation surfaces is provided in this paper, together with 
the (necessary to the proof) extension of a theorem on nonsmooth elastic potential functions. While the former of these generalizations 
is crucial for yield and phase-transformation condition, the latter may find applications for {\it potential energy functions} describing 
phase-transforming
materials, or materials with discontinuous locking in tension, or contact of a body with a discrete elastic/frictional support.
\end{abstract}


\newpage

%\tableofcontents

\section{Introduction}

\subsection*{$\bullet$ Yield or phase-transformation functions}

Bigoni and Piccolroaz (2004) have proposed a new yield (or phase-transformation) function within the class of {\it isotropic} functions of the stress tensor $\bsigma$ defined by
\beq
\lb{funzionazza}	
F(\bsigma) = f(p) + \frac{q}{g(\theta)},
\eeq
in which, having defined 
\beq
\lb{fiegi}
\Phi=\frac{p+c}{p_c+c},
\eeq
the meridian and deviatoric functions take the form
\beq
\lb{effedip}
f(p) =
\left\{
\barr{ll}
-Mp_c\sqrt{\left(\Phi-\Phi^m\right) \left[2(1-\alpha)\Phi+\alpha\right]}, & \Phi \in [0,1], \\[3mm]
+\infty, & \Phi \notin [0,1],
\earr
\right.
\quad
\frac{1}{g(\theta)} = \cos \left[ \beta\frac{\pi}{6} - \frac{\cos^{-1} \left( \gamma \cos 3\theta \right)}{3} \right],
\eeq
respectively, where $p,q$ and $\theta$ are stress invariants.
\footnote{
The stress invariants $p$, $q$ and $\theta$ are defined by 
\beq
p = -\frac{\tr \bsigma}{3}, \quad q = \sqrt{3 J_2}, \quad \theta = \frac{1}{3} \arccos \left( \frac{3\sqrt{3}}{2} \frac{J_3}{J_2^{3/2}} \right),
\eeq
where $J_2$ and $J_3$ the second and third invariant of the deviatoric stress $\bS$
\beq
J_2 = \frac{1}{2} \tr \bS^2, \quad J_3 = \frac{1}{3} \tr \bS^3, \quad \bS = \bsigma - \frac{\tr\bsigma}{3} \bI,
\eeq
in which $\bI$ is the identity tensor.
}

To preserve convexity of the yield surface, the seven material parameters defining the meridian shape function $f(p)$ and the deviatoric shape 
function $g(\theta)$ are restricted to range within the following 
intervals
\beq
M > 0, \quad p_c > 0, \quad c \geq 0, \quad 0 < \alpha < 2, \quad m > 1, \quad
0 \leq \beta \leq 2, \quad 0 \leq \gamma \leq 1.
\eeq

The interest in the above yield function and in the more general class of functions (\ref{funzionazza}) lies in the fact that they can model the 
behaviour of many materials of engineering importance, such as ceramic (Piccolroaz et al.\ 2006) and metal (Bier and Hartmann, 2006; 
Hartmann and Bier, 2008; Heisserer et al.\ 2008) powders, metals (Hu and Wang (2005); Wierzbicki et al.\ 2005; Coppola and Folgarait, 2007), high 
strength alloys (for instance, Inconnel 718) (Bai and Wierzbicki, 2008), shape memory alloys (for instance, NiTi, NiAl, CuZnGa, or CuAlNi) 
(Raniecki and Mr\'{o}z, 2008), concrete (Babua et al.\ 2005), and geomaterials (Dal Maso et al.\ 2007; Descamps and Tshibangu, 2007; DorMohammadi 
and Khoei 2008; Maiolino, 2005; Mortara, 2008; Sheldon et al.\ 2008). Moreover, eqn.\ (\ref{funzionazza}) can be used as a general expression to set 
the condition for phase-transformations, for instance, to determine the stress threshold for martensitic or austenitic transformation (Raniecki 
and Lexcellent, 1998 and Lexcellent et al.\ 2002).

\subsection*{$\bullet$ Convexity of yield or phase-transformation functions}

\begin{figure}[!htb]
\begin{center}
\vspace*{3mm}
(a) \includegraphics[width=7.3cm]{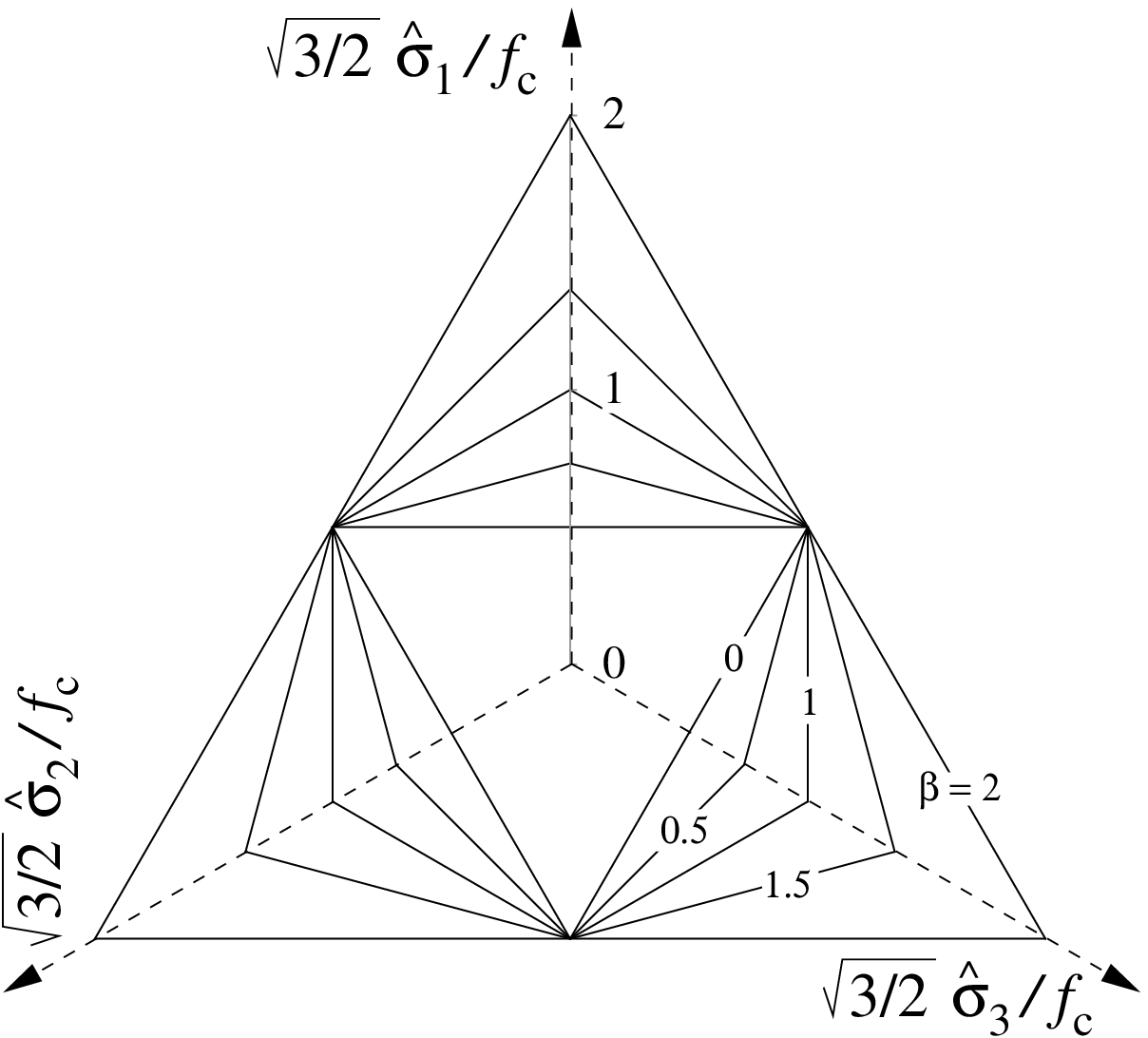}
\hspace{0mm}
(b) \includegraphics[width=6cm]{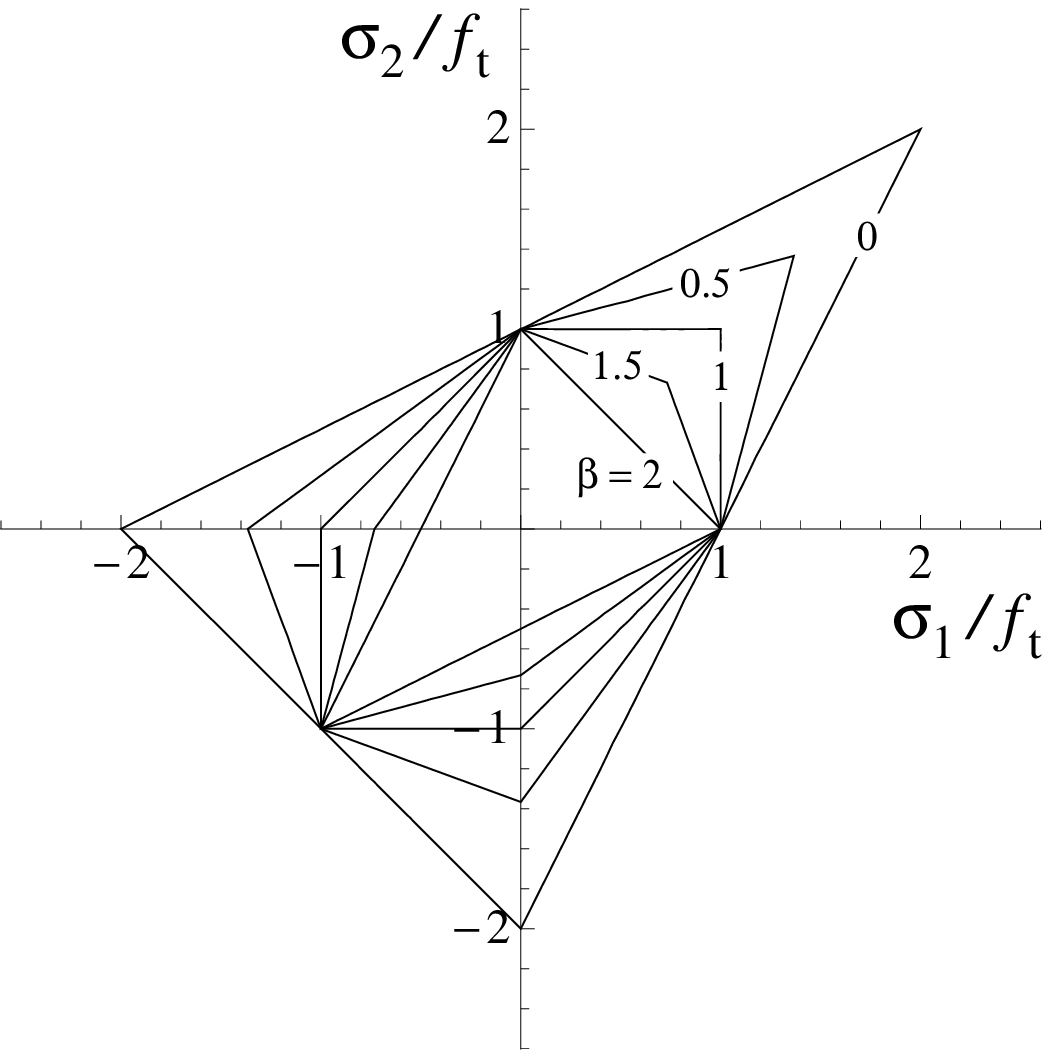}
\caption{\footnotesize 
Examples of yield surfaces with corners obtained within the general class of yield function \eq{funzionazza}, namely 
$F(\bsigma) = -f_c/g(\pi/3) + q/g(\theta)$ with the Bigoni and Piccolroaz (2004) deviatoric function, eqn.\ \eq{effedip}$_2$, taking $\gamma = 1$.
(a) Deviatoric section: $\beta = 0,0.5,1,1.5,2$. 
(b) Yield surface in the biaxial plane $\sigma_1/f_t$ vs. $\sigma_2/f_t$, with $\sigma_3 = 0$: $\beta = 0,0.5,1,1.5,2$; $f_t$ and $f_c$ denote 
tensile and compressive uniaxial yield stress, respectively.}
\label{fig01a}
\end{center}
\end{figure}

With reference to the class of functions (\ref{funzionazza}), Bigoni and Piccolroaz (2004) have proved a general proposition providing necessary 
and sufficient conditions relating convexity of the yield function to convexity of the corresponding yield surface in the Haigh-Westergaard 
stress space (or principal stresses representation),
\footnote{
Four years later, exactly the same proof has been independently published by Raniecki and Mr\'{o}z (2008).
} 
a crucial property in the development of new expressions for yield or phase-transformation criteria. This proof is based on both the hypotheses 
of smoothness of the function $g(\theta)$, 
\footnote{
The function $F(\bsigma)$ given by eqn.\ (\ref{funzionazza}) is always nonsmooth along the hydrostatic axis. However, this fact has no consequences 
on convexity, as shown in Lemma \ref{supersega}. The fact that $f'(p)$ blows up to infinity when $p$ tends to $p_c$ and $-c$, eqn.\ 
(\ref{effedip})$_1$, is the only possibility to obtain smooth closures at the hydrostatic axis.
} 
and validity of the smoothness limiting conditions $g'(0) = 0$ and $g'(\pi/3) = 0$, while as noticed by Laydi and Lexcellent (2009, see Appendix 
\ref{lexcellent} for a detailed discussion), the class of functions (\ref{funzionazza}) --even under the particularizations (\ref{effedip})-- 
may describe deviatoric yield surfaces with corners (Fig.\ \ref{fig01a}), in which case the conditions for convexity provided by Bigoni and Piccolroaz (2004) 
remain only necessary but not sufficient.
\footnote{
It should be noted from Fig.\ \ref{fig01a} that, although the function $g(\theta)$ is smooth, the limiting conditions are $g'(0) < 0 $ and 
$g'(\pi/3) > 0$, so that there are corners (yet the yield surface still results convex, see Theorem \ref{cornerone}). 
}
Since the convexity proposition is fundamental in developing new yield or phase-transformation criteria, it has immediately attracted a 
strong attention (Taillard et al. 2007; Laydi and Lexcellent, 2009; Lavernhe-Taillard et al. 2009; Saint-Sulpice et al. 2009; 
Valoroso and Rosati, 2009) and may definitely be important in analysing yield criteria with corners. Therefore, it becomes imperative to 
generalize the convexity proposition to nonsmooth deviatoric yield surfaces, which is obtained in the present article 
(Theorem \ref{generalissimofranco}, Section \ref{cornerazzi}). 
%%%%%%%%%%%%%%%%%%%%%%%%%%%%%%%%%%%%%%%%%%%%%%%%%%%%%%%%%%%%%%%%%%%%%%
%%%%%%%%%%%%%%%%%%%%%%%%%%%%%%%%%%%%%%%%%%%%%%%%%%%%%%%%%%%%%%%%%%%%%%

\begin{figure}[!htb]
\begin{center}
\vspace*{3mm}
\includegraphics[width=10cm]{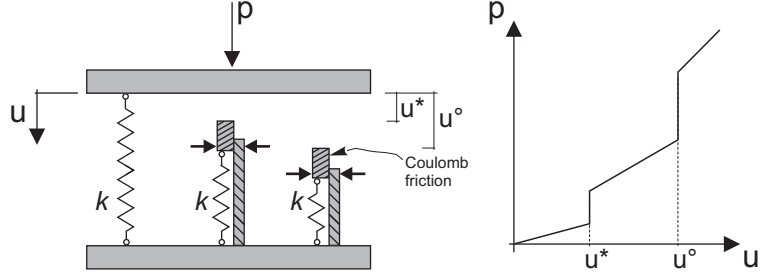}
\caption{\footnotesize 
Contact with a nonlinear constraint. For $u<u^*$ the behaviour is linear elastic, while at $u=u^*$ and $u=u^\circ$ stiffness becomes 
infinite due to contact with frictional asperities obeying a rigid-plastic Coulomb rule. 
Although the sketched model is intrinsically inelastic, a representation restricted
to the loading branch could be described by a continuous, convex and nonsmooth potential, for which Theorem \ref{hillhill} applies.
In tension, a behaviour similar to that sketched on the right would correspond to a discontinuous locking mechanism of a material.}
\label{fig01c}
\end{center}
\end{figure}

The generalization of the Bigoni and Piccolroaz (2004) proposition requires the generalization to nonregular functions of a theorem given by Hill (1968) 
regarding
convexity of elastic strain potentials. In particular, Hill (1968) has shown that convexity 
of a smooth scalar isotropic function of a second-order symmetric tensor (a work-conjugate strain measure in his case) is equivalent 
to the convexity of the corresponding function of the principal values (the principal stretches in his case). 
The Hill's theorem is of fundamental importance, since in many cases [for instance for the Ogden (1982) constitutive equations for rubber 
elasticity and the so-called \lq J$_2$--deformation 
theory materials', Neale (1981)] constitutive equations of finitely-strained elastic
materials are formulated with reference to the principal stretches and not with reference to the tensorial quantities, so that this theorem 
is usually reported in books (see for instance Ogden, 1984). 
Bigoni and Piccolroaz (2004) have recognized that the Hill's theorem can be useful also for yield functions in plasticity theory, 
indeed the theorem has been duplicated (with a slightly different proof, without mentioning Hill's theorem) in the context of elastoplasticity
by Yang (1980). 
However, until now no generalization of the Hill's theorem to nonregular functions has ever been given. Such a generalization may be 
relevant for elastic strain energy functions describing phase-transformation materials, or for elastic potential functions describing contact 
with discrete elastic asperities, or materials with discontinuous locking in tension (Fig.\ \ref{fig01c}), but it is certainly of great 
interest for yield functions, which are often nonsmooth
[for instance Hill (1950), Tresca and Coulomb-Mohr]. The generalization is provided in Section \ref{hillazzo} and is the basis for the 
subsequent generalization of the Bigoni and Piccolroaz (2004) proposition to yield criteria (or transforming functions) with corners 
(Section \ref{cornerazzi}). 
%%%%%%%%%%%%%%%%%%%%%%%%%%%%%%%%%%%%%%%%%%%%%%%%%%%%%%%%%%%%%%%%%%%%%%
%%%%%%%%%%%%%%%%%%%%%%%%%%%%%%%%%%%%%%%%%%%%%%%%%%%%%%%%%%%%%%%%%%%%%%

\section{On smoothness of yield (or phase-transformation) functions}

The conditions for smoothness of function $F(\bsigma)$, eqn.\ (\ref{funzionazza}), can be obtained by analysing the gradient of $F(\bsigma)$, 
\beq
\deriv{F}{\bsigma} = -\frac{1}{3} f'(p) \bI + \sqrt{\frac{3}{2}}\frac{1}{g} \tilde{\bS} - \sqrt{\frac{3}{2}}\frac{g'}{g^2} \tilde{\bS}^\perp,
\eeq
where $\bI$ is the identity tensor and
\beq
\lb{oops}
\tilde{\bS} = \sqrt{\frac{3}{2}}\frac{\bS}{q}, \quad 
\tilde{\bS}^\perp = \sqrt{\frac{2}{3}} q \deriv{\theta}{\bsigma} = 
-\frac{3\sqrt{3}}{\sqrt{2} q^2 \sin 3\theta} \left[ \bS^2 - \frac{2}{9} q^2 \bI - \frac{q}{3} \cos 3\theta \bS \right].
\eeq
Note that:
\begin{itemize}

\item $\tilde{\bS}$ is discontinuous along the hydrostatic axis, but this discontinuity does not affect convexity of $F(\bsigma)$, see 
Lemma \ref{supersega};

\item $\tilde{\bS}^\perp$ is discontinuous along the hyperplanes defined by $\theta = 0$ and $\theta = \pi/3$. This discontinuity 
can be eliminated for functions $g$ such that $g'= 0$ at $\theta = 0$ and $\theta = \pi/3$, which is the case of many yield functions, 
for instance, all yield functions described by eqn.\ (\ref{effedip})$_2$ when $0 \leq \gamma < 1$. The analysis of the 
nonsmooth case, at $\theta = 0$ and $\theta = \pi/3$
is the main target of the present article and leads to Theorem \ref{cornerone}, which is generalized into Theorem \ref{supercorner} and 
finally leads to Theorem \ref{generalissimofranco}.

\end{itemize}

We analyse now smoothness of the deviatoric part $q/g(\theta)$ as a function of $S_1,S_2$, where $S_1,S_2$ denote two principal values of the 
deviatoric stress. Assuming that $g(\theta)$ is continuous and strictly positive in $[0,\pi/3]$, and smooth everywhere in $(0,\pi/3)$, the 
gradient of $q/g(\theta)$ with respect to the variables $S_1,S_2$ 
is given by
\beq
\deriv{q/g(\theta)}{S_i} = \frac{1}{g(\theta)}\deriv{q}{S_i} - q \frac{g'(\theta)}{g^2(\theta)} \deriv{\theta}{S_i}, \qquad i = 1,2,
\eeq
where \footnote{
Note that there is a misprint in (Bigoni and Piccolroaz, 2004): their eqn.\ (39)$_1$ should be replaced by eqn.\ (\ref{aringa}).
}
\beq
\lb{aringa}
\deriv{q}{S_i} = \frac{3}{2q}[2S_i - (-1)^i m_i], \qquad i = 1,2, 
\eeq
\beq
\deriv{\theta}{S_i} = - \frac{3\sqrt{3}}{2q^2} \hat{H}(S_1,S_2) m_i, \qquad i = 1,2,
\eeq
in which the indices are not summed and the vector $\bm$ has the components: $\{ \bm \} = \{ S_2, -S_1 \}$.

\begin{figure}[!htb]
\begin{center}
\vspace*{3mm}
\includegraphics[width=6cm]{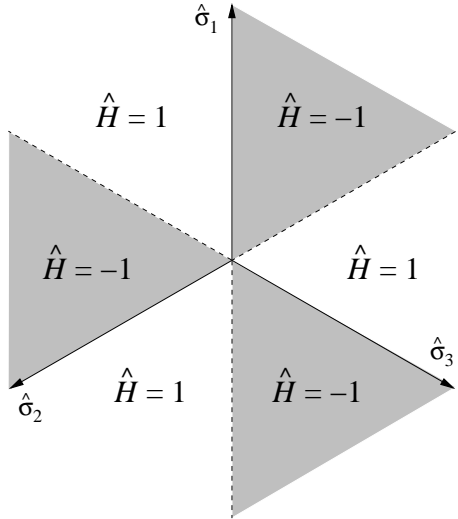}
\caption{\footnotesize 
Plot of the function $\hat{H}$, eqn.\ (\ref{accazzo}), in the deviatoric plane. Axes $\hat{\sigma}_1,\hat{\sigma}_2,\hat{\sigma}_3$ are the projections of the principal stress 
axes onto the deviatoric plane.}
\label{figA}
\end{center}
\end{figure}

The function $\hat{H}(S_1,S_2)$ is a piecewise constant function defined by 
\beq
\lb{accazzo}
\hat{H}(S_1,S_2) = \frac{(S_1 - S_2)(2S_1 + S_2)(S_1 + 2S_2)}{\sqrt{(S_1 - S_2)^2(2S_1 + S_2)^2(S_1 + 2S_2)^2}} = 
\sign[(S_1 - S_2)(2S_1 + S_2)(S_1 + 2S_2)],
\eeq
which takes the values $1$ or $-1$ only, Fig.\ \ref{figA}. 
%%%%%%%%%%%%%%%%%%%%%%%%%%%%%%%%%%%%%%%%%%%%%%%%%%%%%%%%%%%%%%%%%%%%%%
%%%%%%%%%%%%%%%%%%%%%%%%%%%%%%%%%%%%%%%%%%%%%%%%%%%%%%%%%%%%%%%%%%%%%%
We may note from Fig.\ \ref{figA} that the function $\hat{H}(S_1,S_2)$ is discontinuous along the projections of the principal stress axes on the deviatoric 
plane:
$$
\text{axis } \hat{\sigma}_1: \{S_1, -S_1/2, -S_1/2\}, \quad \text{if } S_1 > 0 \text{ then } \theta = 0, \quad 
\text{if } S_1 < 0 \text{ then } \theta = \pi/3, 
$$
\beq
\lb{assicazzi}
\text{axis } \hat{\sigma}_2: \{-S_2/2, S_2, -S_2/2\}, \quad \text{if } S_2 > 0 \text{ then } \theta = 0, \quad 
\text{if } S_2 < 0 \text{ then } \theta = \pi/3, 
\eeq
$$
\text{axis } \hat{\sigma}_3: \{-S_3/2, -S_3/2, S_3\}, \quad \text{if } S_3 > 0 \text{ then } \theta = 0, \quad 
\text{if } S_3 < 0 \text{ then } \theta = \pi/3. 
$$

\begin{figure}[!htb]
\begin{center}
\vspace*{3mm}
(a) \includegraphics[width=6cm]{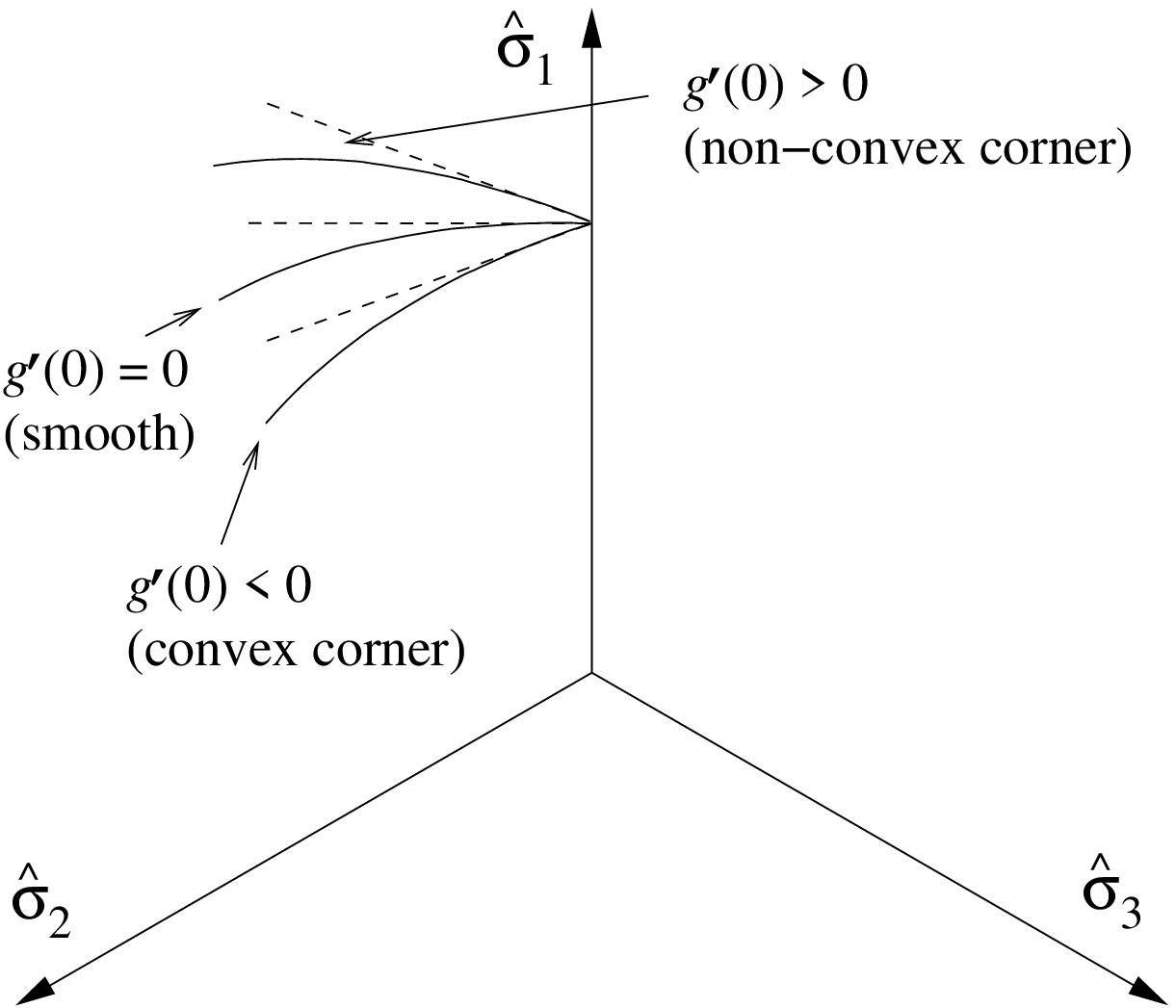}
\hspace{5mm}
(b) \includegraphics[width=6cm]{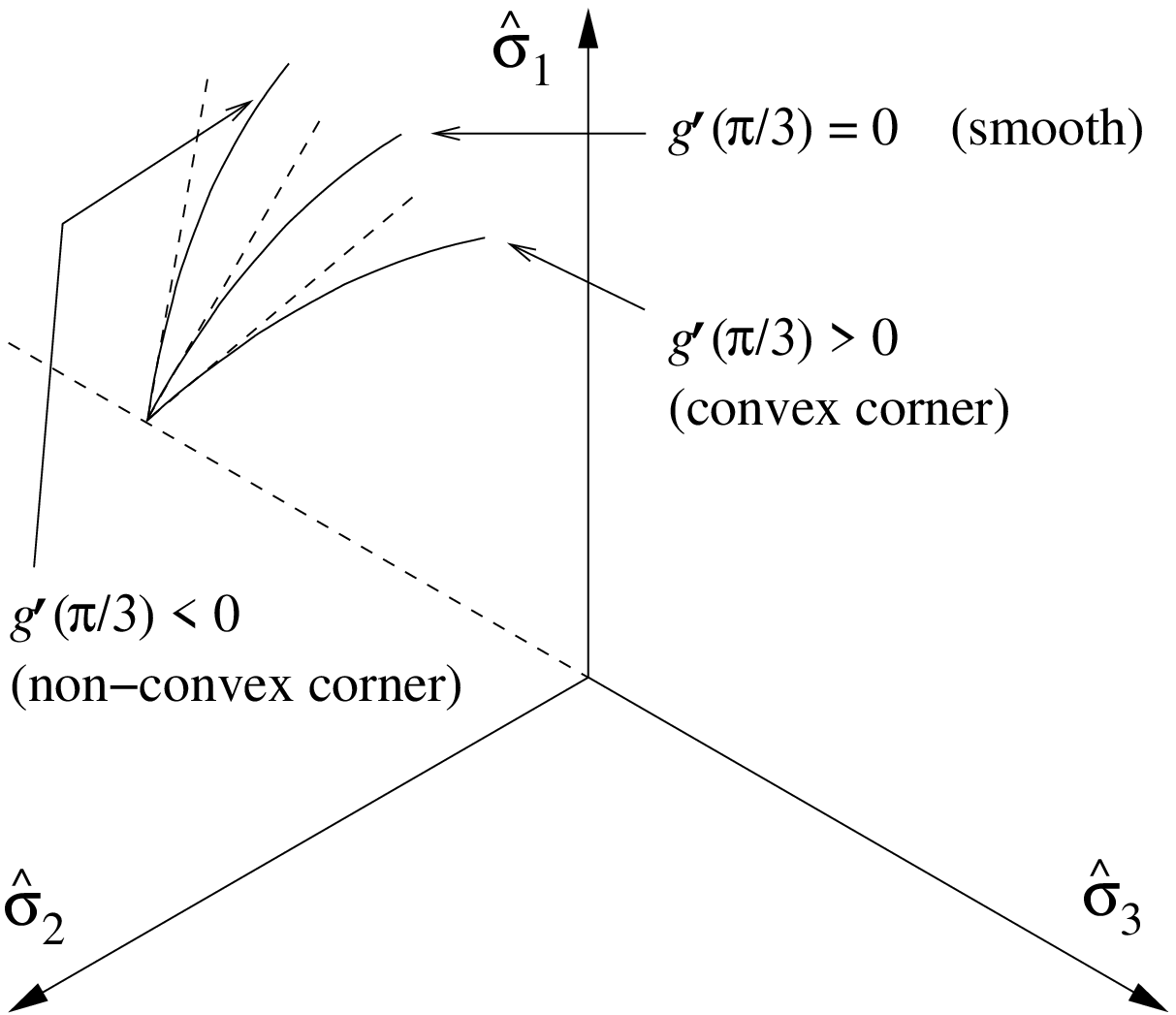}
\caption{\footnotesize 
Conditions for smoothness of the yield surface deviatoric section. (a) At $\theta = 0$: $g'(0) = 0$. (b) At $\theta = \pi/3$: $g'(\pi/3) = 0$.}
\label{figB}
\end{center}
\end{figure}

Accordingly, 
\begin{quote}
{\it the function $q/g(\theta)$ is smooth if and only if $g(\theta)$ is smooth everywhere in $(0,\pi/3)$ and $g'(0) = g'(\pi/3) = 0$}, see 
Fig.\ \ref{figB}.
\end{quote}

%%%%%%%%%%%%%%%%%%%%%%%%%%%%%%%%%%%%%%%%%%%%%%%%%%%%%%%%%%%%%%%%%%%%%%
%%%%%%%%%%%%%%%%%%%%%%%%%%%%%%%%%%%%%%%%%%%%%%%%%%%%%%%%%%%%%%%%%%%%%%

\section{Nonsmooth, convex and isotropic functions}
\lb{hillazzo}

With reference to elastic potential of finite-strain constitutive equations, Hill (1968) has proven that convexity 
of a smooth scalar isotropic function of a second-order symmetric tensor (a work-conjugate strain measure in his case) is equivalent 
to the convexity of the corresponding function of the principal values (the principal stretches in his case). 
The Hill's theorem is of fundamental importance, since in many cases constitutive equations of finitely-strained elastic
materials are formulated with reference to the principal stretches and not with reference to the tensorial quantities, 
see for instance Ogden (1984). 
Bigoni and Piccolroaz (2004) have evidenced that the Hill's theorem also applies to yield functions in elastoplasticity theory. 

Until now, no generalization of the Hill's theorem to nonsmooth function has ever been given. As mentioned in the Introduction, 
such a generalization may be relevant for elastic strain energy functions describing phase-transformation materials (although in those cases 
usually non-convexity is employed), or for potential functions describing discontinuous locking in tension, or contact with discrete elastic 
springs (as explained in 
Fig.\ \ref{fig01c}, where contact of a rigid punch with a linear elastic set of springs having different heights and rigid/frictional devices is 
envisaged. The loading branch of this model can be described through a piecewise linear and convex strain energy function).

In any case, the generalization of the Hill's theorem is certainly of great 
interest for yield functions, which are often nonsmooth, as for instance in the cases of 
the Hill (1950), Tresca, modified-Tresca, and Coulomb-Mohr yield surfaces.
The generalization is provided in this Section. 

We begin with a simple lemma. 

\begin{lemma}
Let us consider a scalar isotropic function $\phi$ of tensorial argument $\sigma_{ij} \in \Sym$ and 
the corresponding function $\tilde{\phi}$ written with reference to the principal values $\sigma_i$:
$$
\phi(\sigma_{11}, \sigma_{22}, \sigma_{33}, \sigma_{12}, \sigma_{13}, \sigma_{23}) = \tilde{\phi}(\sigma_1,\sigma_2,\sigma_3),
$$ 
then, due to isotropy, the following equality holds
\beq
\lb{orestebursi}
\tilde{\phi}(\sigma_1,\sigma_2,\sigma_3) = \phi(\sigma_{1}, \sigma_{2}, \sigma_{3}, 0, 0, 0),
\eeq
so that $\tilde{\phi}$ is the restriction of $\phi$ to the subdomain of diagonal tensors.
\end{lemma}

\begin{proof}
The property \eq{orestebursi} is easily proven by the following consideration. The isotropy of $\phi(\bsigma)$ implies that the function 
$\phi(\bsigma)$ is equal to a function $\hat{\phi}$ of the invariants of $\bsigma$,
$$
\phi(\sigma_{11}, \sigma_{22}, \sigma_{33}, \sigma_{12}, \sigma_{13}, \sigma_{23}) = \hat{\phi}(\tr \bsigma,\tr \bsigma^2, \tr \bsigma^3),
$$
and thus
$$
\tilde{\phi}(\sigma_1,\sigma_2,\sigma_3) = \hat{\phi}(\tr \bsigma,\tr \bsigma^2, \tr \bsigma^3) = \phi(\sigma_{1}, \sigma_{2}, \sigma_{3}, 0, 0, 0).
$$
\end{proof}

We need now to introduce the notion of {\em subdifferential} (or {\em subgradient}), which will be used in the sequel for the generalization of the Hill (1968) 
theorem to nonsmooth functions.

A function $\phi: \, U \subseteq \Reals^n \rightarrow \Reals$ is convex if and only if the subdifferential 
\beq
\partial \phi(\bX_0) = \left\{\bQ \in \Reals^n: \, \phi(\bX) -\phi(\bX_0) \geq \bQ \scalp (\bX-\bX_0), \quad \forall \bX \in U \right\},
\eeq
is defined and non empty  at every point $\bX_0$ of its domain $U$.

Note that although the subgradient is a  set of vectors, in the sequel we shall denote with the term \lq subgradient' both the set itself and 
its elements.

The following lemma, necessary to the proof of Theorem (\ref{hillhill}), is similar to the analogous given by Hill (1968), but now
it has been generalized and extended to nonsmooth isotropic functions.

\begin{lemma}
\lb{ordinenuovo} 
Given a convex function of the principal stresses, $\tilde{\phi}(\sigma_1,\sigma_2,\sigma_3)$, the algebraic order of components of the subgradient $(Q_1,Q_2,Q_3)$ at $(\sigma_1,\sigma_2,\sigma_3)$ is the same as $(\sigma_1,\sigma_2,\sigma_3)$.
\end{lemma}

\begin{proof}
From the strict convexity of $\tilde{\phi}$, it follows that 
\beq
\sum_{i=1}^3(Q_i - Q_i^0) (\sigma_i - \sigma_i^0) > 0, 
\eeq
$\forall (Q_1,Q_2,Q_3) \in \partial \tilde{\phi}(\sigma_1,\sigma_2,\sigma_3)$ and 
$\forall (Q_1^0,Q_2^0,Q_3^0) \in \partial \tilde{\phi}(\sigma_1^0,\sigma_2^0,\sigma_3^0)$.
Choosing $(\sigma_2^0,\sigma_1^0,\sigma_3^0) = (\sigma_1,\sigma_2,\sigma_3)$ and {\em taking into account isotropy}, 
it follows that
\beq
(Q_1 - Q_2)(\sigma_1 - \sigma_2) > 0,
\eeq
and similarly for each of the other pairs. It follows that the vector $(Q_1,Q_2,Q_3)$ is ordered in the same 
algebraic order as $(\sigma_1,\sigma_2,\sigma_3)$, a property which remains true also assuming convexity \lq $\geq$' instead of 
strict convexity \lq $>$'.
\end{proof}

Note also that we will make use in the proof of Theorem \ref{hillhill} of an auxiliary property of the scalar product between two symmetric 
tensors, first noticed by Hill (1968):
\begin{quote}
\em
``if their eigenvalues are given, but their axes are directly arbitrarily, the product attains its greatest value when the major and minor axes are 
pairwise coincident.''
\end{quote}

We refer to Appendix \ref{theproof} for a detailed discussion and proof of this auxiliary property.

\begin{theorem} 
\lb{hillhill}
\underline{Extension of the Hill (1968) theorem to nonregular functions}. Convexity of an isotropic (not necessarily smooth) function of a symmetric (stress) tensor 
$\bsigma$ is equivalent to convexity of the corresponding function of the principal (stress) values $\sigma_i$ ($i=1,2,3$).
In symbols, given:
\beq
\phi(\bsigma) = \tilde{\phi}(\sigma_1, \sigma_2, \sigma_3),
\eeq
then $\forall \bsigma \in \Sym$, 
\beq
\lb{asterix}
\exists \bQ \in \Sym :\ \phi(\bsigma') - \phi(\bsigma) \geq \bQ \cdot \left(\bsigma' - \bsigma\right), \quad \forall \bsigma' \in \Sym, 
\eeq
$$
\Updownarrow
$$
\beq
\lb{obelix}
\exists (Q_1,Q_2,Q_3) \in \Reals^3 :\ \tilde{\phi}(\sigma_1',\sigma_2',\sigma_3') - \tilde{\phi}(\sigma_1,\sigma_2,\sigma_3) \geq 
\sum_{i=1}^3 Q_i \left(\sigma_i' - \sigma_i\right), \quad \forall (\sigma_1',\sigma_2',\sigma_3') \in \Reals^3.
\eeq
\end{theorem}

\begin{proof}
The proof that (\ref{asterix}) $\Longrightarrow$ (\ref{obelix}) follows immediately from the property (\ref{orestebursi}). The
converse (\ref{obelix}) $\Longrightarrow$ (\ref{asterix}) is not trivial and is proven in the following.

We denote by $(\sigma_1,\sigma_2,\sigma_3)$ the principal values of a given $\bsigma$ and by $(Q_1,Q_2,Q_3)$ the subgradient of 
$\tilde{\phi}$ at $(\sigma_1,\sigma_2,\sigma_3)$. We define now $\bQ \in \Sym$ to be 
$$
\bQ = Q_1 \bq_1 \otimes \bq_1 + Q_2 \bq_2 \otimes \bq_2 + Q_3 \bq_3 \otimes \bq_3,
$$ 
where $\{\bq_1,\bq_2,\bq_3\}$ is an orthonormal basis of $\Reals^3$. Then, assuming that $(\sigma_1',\sigma_2',\sigma_3')$ are numbered in the same 
algebraic order as $(\sigma_1,\sigma_2,\sigma_3)$, and since from the Lemma \ref{ordinenuovo} we know that the algebraic order of 
$(Q_1,Q_2,Q_3)$ is also the same as $(\sigma_1,\sigma_2,\sigma_3)$, the auxiliary property of the scalar product (proven in Appendix \ref{theproof}), 
implies that
\beq
\lb{diseq}
\sum_{i=1}^3 Q_i \left(\sigma_i' - \sigma_i\right) \geq \bQ \cdot (\bsigma' - \bsigma), \quad \forall \bsigma' \in \Sym.
\eeq

Since, by hypothesis, the following equation holds true
\beq
\lb{hypo}
\phi(\bsigma') - \phi(\bsigma) = \tilde{\phi}(\sigma_1',\sigma_2',\sigma_3') - \tilde{\phi}(\sigma_1,\sigma_2,\sigma_3) \geq 
\sum_{i=1}^3 Q_i \left(\sigma_i' - \sigma_i\right), \quad \forall (\sigma_1',\sigma_2',\sigma_3') \in \Reals^3,
\eeq
eqn.\ \eq{diseq} guarantees that $\bQ \in \partial \phi(\bsigma)$, so that $\phi(\bsigma)$ results to be convex.
\end{proof}

\section{Convexity of yield functions with corners}
\lb{cornerazzi}

We begin with proving that the discontinuity of the yield function gradient along the hydrostatic axis [see eqn.\ \eq{oops}] is inconsequential on convexity.

\begin{lemma}\lb{supersega} 
The convexity of the function $q/g(\theta)$ is unaffected by the fact that $\tilde{\bS}$ and $\tilde{\bS}^\perp$ defined in eqn.\ \eq{oops} are discontinuous along 
the hydrostatic axis, where $S_1 = S_2 = 0$.
\end{lemma}

\begin{proof}
From the definition of convexity, it follows that (van Tiel, 1984)
\begin{quote}
$\ds q/g(\theta)$ is convex at $(S_1,S_2)$ $\Leftrightarrow$ 
\begin{minipage}[t]{8cm}
for every line $t$ through $(S_1,S_2)$, the restriction to $t$ of $\ds q/g(\theta)$ is convex
 \end{minipage}
\end{quote}

Let us consider all deviatoric lines through the point $\{S_1=0, S_2=0\}$, these can be represented (using parameter $\epsilon$ and slope $k$) as
$\{\epsilon, k\, \epsilon\}$. The restriction of $q/g(\theta)$ to these lines is a function $h(\epsilon)$, whose derivative with respect to 
$\epsilon$ is
\beq
\lb{derivatazza}
h'(\epsilon) = \left\{ \frac{\nabla q}{g}-q \frac{g'}{g^2}\nabla \theta \right\} \scalp \{ 1, k \},
\eeq
where $\nabla$ denotes the gradient taken with respect to the variables $\{S_1, S_2\}$, so that, since 
$\nabla \theta \scalp \{1, k\} = 0$ and 
\beq
q = |\epsilon| \sqrt{3} \sqrt{1+k+k^2}, \quad \nabla q = \frac{\sqrt{3}\, \sign{\epsilon}}{\sqrt{1+k+k^2}}\{2+k,1+2k\},
\eeq
we obtain 
\beq
h'(\epsilon) = \frac{\sqrt{3}\, \sign{\epsilon}}{g} \sqrt{1+k+k^2}.
\eeq
At a singular point, convexity requires that 
\beq
\lim_{\epsilon \rightarrow 0^-} h'(\epsilon) < \lim_{\epsilon \rightarrow 0^+} h'(\epsilon),
\eeq
which is always satisfied, so that the discontinuities in $\tilde{\bS}$  and  $\tilde{\bS}^\perp$ along the hydrostatic axis are inconsequential 
on convexity.
\end{proof}

With reference to the deviatoric part $q/g(\theta)$ of the yield function (\ref{funzionazza}), we give now necessary and sufficient conditions 
(Theorem \ref{cornerone}) for equivalence between convexity of yield functions and convexity of yield surface. To this purpose, we first need 
the following lemma.

\begin{lemma}
\lb{lemmino}
Given a generic isotropic function $\phi$ of the stress that can be expressed as
\beq
\tilde{\phi}(\sigma_1, \sigma_2, \sigma_3) = \hat{\phi}(S_1,S_2),
\eeq
where $S_1$ and $S_2$ are two of the principal components of the deviatoric stress, i.e.
\beq
\lb{devia}
S_1= \frac{1}{3} \left(2\sigma_1-\sigma_2-\sigma_3\right), \quad
S_2= \frac{1}{3} \left(-\sigma_1+2\sigma_2-\sigma_3\right),
\eeq
convexity of $\tilde{\phi}(\sigma_1, \sigma_2, \sigma_3)$ is equivalent to convexity of $\hat{\phi}(S_1,S_2)$.
\end{lemma}

\begin{proof}
This proposition follows immediately from the fact that the relation (\ref{devia}) between $\{S_1,S_2\}$ and $\{\sigma_1, \sigma_2, \sigma_3\}$ is linear. 
\end{proof}

The following theorem is the generalization of Lemma 3 by Bigoni and Piccolroaz (2004) to the case of
nonsmooth deviatoric sections of the yield surface. Note that the difference between the two versions of the theorem lies on 
the two conditions $g'(0) \leq 0$ and $g'(\pi/3) \ge 0$. A consequence of the following theorem 
is that the convexity conditions by Laydi and Lexcellent (2009) are only sufficient (but not necessary) for convexity of smooth functions 
(see Appendix \ref{lexcellent}).

\begin{theorem} 
\lb{cornerone} 
\underline{Convexity of nonsmooth deviatoric representation $q/g(\theta)$ vs. convexity of the deviatoric} 
\underline{section of the yield  surface.}

Assuming that $g(\theta)$ is continuous and strictly positive in $[0,\pi/3]$ and twice-differentiable everywhere in $(0,\pi/3)$, convexity of 
\beq
\lb{roteta}
\frac{q}{g(\theta)}
\eeq
as a function of $S_1, S_2$ is equivalent to the convexity of the deviatoric section in the Haigh-Westergaard space:
\beq
\lb{curvatura}
g^2 + 2g'^2 - gg'' \geq 0, \quad \forall \theta \in (0,\pi/3) \quad \text{and} \quad g'(0) \leq 0, \quad g'(\pi/3) \ge 0.
\eeq
\end{theorem}

\begin{proof}

Let us define $\Omega$ as the set of all points $\{S_1,S_2\}$ not on the axes $\hat{\sigma}_1,\hat{\sigma}_2,\hat{\sigma}_3$, 
see eqns.\ (\ref{assicazzi}). The theorem is proven first (point 1 below) by showing local convexity at all points of $\Omega$ (regular points) and,
second (point 2 below), considering the points of $\partial\Omega$, i.e. the axes $\hat{\sigma}_1,\hat{\sigma}_2,\hat{\sigma}_3$, where the 
function has corners (singular points).

\bit

\item[1)] Local convexity in $\Omega$ (for which $0 < \theta < \pi/3$).

The function $q/g(\theta)$ is $C^2(\Omega)$, so that we can apply the convexity criterion based on the Hessian.

The Hessian of the function \eq{roteta} is
\beq
\lb{iniz}
\frac{\partial^2 q/g(\theta)}{\partial S_i \partial S_j} = 
\frac{1}{g^3} \left[g^2 \frac{\partial^2 q}{\partial S_i \partial S_j} 
+ q (2g'\,^2 - gg'') \frac{\partial \theta}{\partial S_i} \frac{\partial \theta}{\partial S_j} 
- gg' \left(\frac{\partial q}{\partial S_i} \frac{\partial \theta}{\partial S_j} 
+ \frac{\partial q}{\partial S_j} \frac{\partial \theta}{\partial S_i} 
+ q \frac{\partial^2 \theta}{\partial S_i \partial S_j}\right)\right],
\eeq
where $i$ and $j$ range between 1 and 2 and all functions $q$ and $\theta$ are to be understood as functions of $S_1$ and $S_2$ only. The Hessian 
of $q$ may be easily calculated to be
$$
\frac{\partial^2 q}{\partial S_i \partial S_j} = \frac{27}{4 q^3} m_i m_j,
$$
where indices are not summed and $\{\bm\} = \{S_2,-S_1\}$. The Hessian of $\theta$ becomes
$$
\frac{\partial^2 \theta}{\partial S_i \partial S_j} = \frac{-1}{3 \sin 3\theta} 
\left(\frac{\cos 3\theta}{\sin^2 3\theta} \frac{\partial \cos 3\theta}{\partial S_i} \frac{\partial \cos 3\theta}{\partial S_j} + 
\frac{\partial^2 \cos 3\theta}{\partial S_i \partial S_j} \right),
$$
so that 
\beq
\lb{pizza}
\frac{\partial q}{\partial S_i} \frac{\partial \theta}{\partial S_j} + \frac{\partial q}{\partial S_j} \frac{\partial \theta}{\partial S_i} + 
q \frac{\partial^2 \theta}{\partial S_i \partial S_j} = 
\frac{-1}{3 \sin 3\theta}
\left[
\frac{\partial^2 q \cos 3\theta}{\partial S_i \partial S_j} - \cos 3\theta \frac{\partial^2 q}{\partial S_i \partial S_j} + 
q \frac{\cos 3\theta}{\sin^2 3\theta} \frac{\partial \cos 3\theta}{\partial S_i} \frac{\partial \cos 3\theta}{\partial S_j}
\right],
\eeq
where\footnote{
Note that there is a misprint in (Bigoni and Piccolroaz, 2004): their eqns.\ (43)$_1$ should be replaced by eqn.\ (\ref{treteta})$_1$.
}
\beq
\lb{treteta}
\frac{\partial \cos 3\theta}{\partial S_i} = \frac{9 \sqrt{3} \sin 3 \theta}{2 q^2} \hat{H}(S_1,S_2) m_i, \quad
\frac{\partial^2 q \cos 3\theta}{\partial S_i \partial S_j} = - 27^2 \frac{J_3}{q^6} m_i m_j.
\eeq

A substitution of (\ref{treteta}) into (\ref{pizza}) yields
\beq
\ds{\frac{\partial q}{\partial S_i} \frac{\partial \theta}{\partial S_j}
+\frac{\partial q}{\partial S_j} \frac{\partial \theta}{\partial S_i}
+q \frac{\partial^2 \theta}{\partial S_i  \partial S_j}}
= 0,
\eeq
so that we may conclude that the Hessian (\ref{iniz}) can be written as
\beq
\lb{hessian}
 \frac{\partial^2 q/g(\theta)}{\partial S_i \partial S_j} 
 = \frac{27}{4}
\frac{\left( g^2 + 2g'\,^2 -g g'' \right)}{q^3 g^3} m_i \, m_j,
\eeq
from which condition $g^2 + 2g'\,^2 -g g'' \geq 0 $ is immediately obtained.
\item[2)] Local convexity on $\partial \Omega$ (for which $\theta = 0$ or $\theta = \pi/3$).

We consider in the following only the axis $\hat{\sigma}_1$, since the proof remains strictly similar for the other axes.

\bit 

\item[2.1)] Case $\theta = 0$.

A line $t$ through $(1,-1/2)$ has the parametric representation $\{(1 + \epsilon, -1/2 + k \epsilon) | \epsilon \in \Reals\}$, where $\epsilon$ is 
the parameter and $k$ is the slope of the line. Using this representation, the restriction to $t$ of $q/g(\theta)$ is a function $h(\epsilon)$, 
whose derivative is given by eqn.\ (\ref{derivatazza}). From the limits
\beq
q \to 3/2, \quad \nabla q \to \{-3/2,0\} \quad \text{and} \quad \nabla \theta \to \pm \frac{1+2k}{|1+2k|\sqrt{3}} \{1,2\}
\quad \text{as} \quad \epsilon \to 0^\pm,
\eeq
we derive
\beq
h'_\pm(0) = \frac{3}{2g} \mp \frac{\sqrt{3}}{2} |1+2k| \frac{g'(0)}{g^2(0)}.
\eeq
so that the convexity condition for $h(\epsilon)$ at $\epsilon = 0$, namely, $h'_-(0) < h'_+(0)$ is equivalent to $g'(0) < 0$.

\item[2.2)] Case $\theta = \pi/3$.

A line $t$ through $(-1,1/2)$ has the representation $\{(-1 + \epsilon, 1/2 + k \epsilon) | \epsilon \in \Reals\}$, where $k$ is the slope of the 
line. Using this representation, the restriction to $t$ of $q/g(\theta)$ is a function $h(\epsilon)$, whose derivative is given by 
eqn.\ (\ref{derivatazza}).
From the limits
\beq
q \to 3/2, \quad \nabla q \to \{-3/2,0\} \quad \text{and} \quad \nabla \theta \to \mp \frac{1+2k}{|1+2k|\sqrt{3}} \{1,2\}  
\quad \text{as} \quad \epsilon \to 0^\pm,
\eeq
we derive
\beq
h'_\pm(0)  = -\frac{3}{2g} \pm \frac{\sqrt{3}}{2} |1+2k| \frac{g'(0)}{g^2(0)},
\eeq
so that the convexity condition for $h(\epsilon)$ at $\epsilon = 0$, $h'_-(0) < h'_+(0)$ is equivalent to $g'(\pi/3) > 0$.

\eit

\eit

Since $q/g(\theta)$ is locally convex in the sets $\Omega$ and $\partial\Omega$, the proof is concluded by noting that 
$\Omega \cup \partial \Omega$ represents the whole deviatoric plane, so that $q/g(\theta)$ is globally convex.

\end{proof}

\begin{figure}[!htb]
\begin{center}
\vspace*{3mm}
(a) \includegraphics[width=6cm]{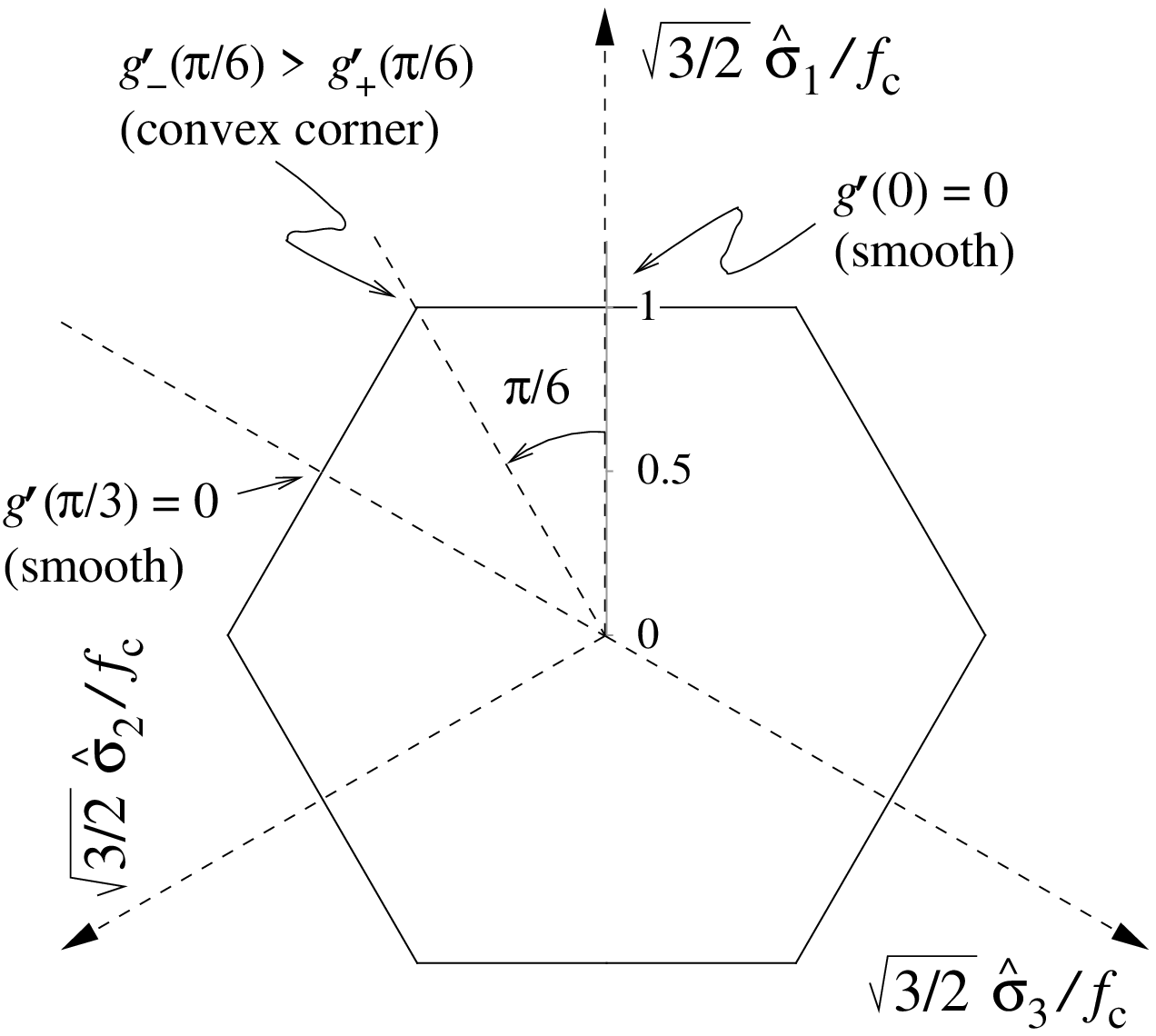}
\hspace{5mm}
(b) \includegraphics[width=6cm]{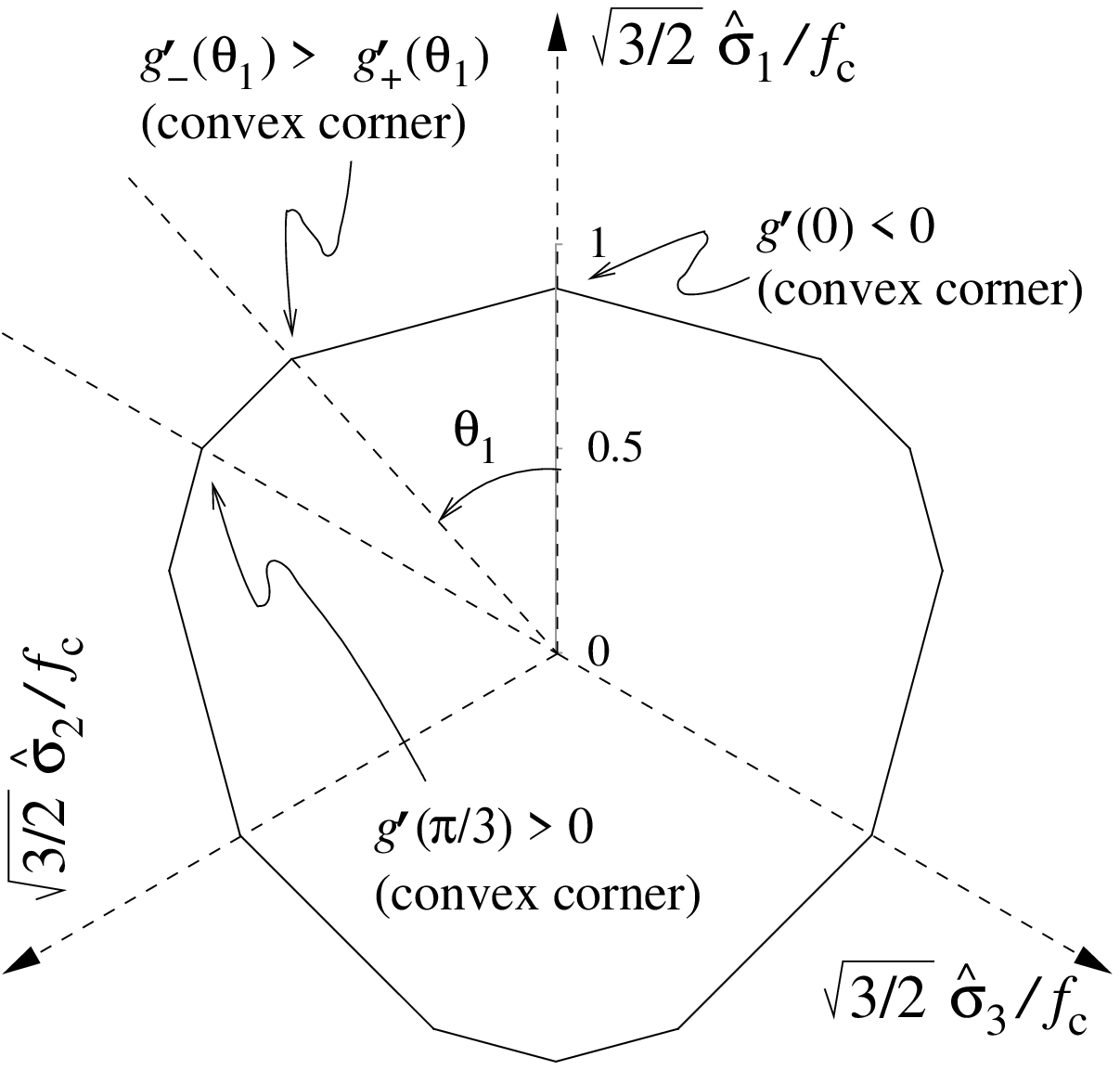}
\caption{\footnotesize 
Yield surface deviatoric sections presenting corners for values of $\theta$ internal to the interval $(0,\pi/3)$: (a) Yield criterion proposed by Hill (1950), described by eqn.\ \eq{wecan}. 
(b) Yield criterion described by eqn.\ \eq{marroni}.}
\label{figC}
\end{center}
\end{figure}

There are yield surfaces, for instance that proposed by Hill (1950), see Fig.\ \ref{figC} (a), presenting corners for values of $\theta$ internal to 
the interval $(0,\pi/3)$. 
In particular, assuming a piecewise smooth function $g(\theta)$, the Hill (1950) criterion can be formulated within the general class 
of yield functions \eq{funzionazza}, namely\footnote{
Bigoni and Piccolroaz (2004) have noted that the Hill criterion cannot be expressed by the function $g(\theta)$, eqn.\ (\ref{effedip})$_2$, defined on the whole interval 
$\theta \in [0, \pi/3]$ through a unique value of parameter $\beta$. 
}, introducing the yield stress in uniaxial compression $-f_c$, by 
$F(\bsigma) = - f_c + q/g(\theta)$, where
\beq
\lb{wecan}
\frac{1}{g(\theta)} = 
\left\{
\barr{ll}
\ds \cos\left[-\frac{1}{3} \cos^{-1}(\cos 3\theta)\right], & \ds 0 \leq \theta \leq \pi/6, \\[5mm]
\ds \cos\left[\frac{\pi}{3} - \frac{1}{3} \cos^{-1}(\cos 3\theta)\right], & \ds \pi/6 < \theta \leq \pi/3.
\earr
\right.
\eeq

Another example of a deviatoric section with corners in $\theta = 0$, $\theta = \pi/3$, and $\theta = \theta_1 = 7\pi/30$ is given by 
$F(\bsigma) = - f_c/g(\pi/3) + q/g(\theta)$, with
\beq
\lb{marroni}
g(\theta) = 
\left\{
\barr{ll}
\ds \frac{\cos\left[\pi/12 - 1/3 \cos^{-1}(\cos 3\theta_1)\right]}{\cos\left[\pi/12 - 1/3 \cos^{-1}(\cos 3\theta)\right]}, 
& 
\ds 0 \leq \theta \leq \theta_1, \\[5mm]
\ds \frac{\cos\left[\pi/4 - 1/3 \cos^{-1}(\cos 3\theta_1)\right]}{\cos\left[\pi/4 - 1/3 \cos^{-1}(\cos 3\theta)\right]}, 
& 
\ds \theta_1 < \theta \leq \pi/3,
\earr
\right.
\eeq
which is plotted in Fig.\ \ref{figC} (b). 
%%%%%%%%%%%%%%%%%%%%%%%%%%%%%%%%%%%%%%%%%%%%%%%%%%%%%%%%%%%%%%%%%%%%%%
%%%%%%%%%%%%%%%%%%%%%%%%%%%%%%%%%%%%%%%%%%%%%%%%%%%%%%%%%%%%%%%%%%%%%%

It is clear from the above examples that employing the function $g(\theta)$ defined by eqn.\ (\ref{effedip})$_2$, with different values of parameter $\beta$
on a finite number of subintervals of $\theta \in [0, \pi/3]$, it is possible to represent all possible nonsmooth deviatoric sections of a yield 
surface. This statement justifies the interest in the following theorem, covering the situations in which the yield surface presents 
corners for values of $\theta$ internal to $(0,\pi/3)$.

\begin{theorem}
\lb{supercorner}  
\underline{Convexity of piecewise-smooth deviatoric representation $g(\theta)$ vs. convexity of the deviatoric} 
\underline{section of the yield  surface.}

Assuming that $g(\theta)$ is continuous and strictly positive in $[0,\pi/3]$ and twice-differentiable almost everywhere in $(0,\pi/3)$, and 
denoting by $\theta_i \in (0, \pi/3)$ the singular points of $g(\theta)$, convexity of 
\beq
\lb{rotetapoi}
\frac{q}{g(\theta)}
\eeq
as a function of $S_1, S_2$ is equivalent to the convexity of the deviatoric section in the Haigh-Westergaard space:
\beq
\lb{curvaturapoi}
g^2 + 2g'^2 - gg'' \geq 0, \quad \forall \theta \in (0,\pi/3)-\{\theta_i\} \quad \text{and} \quad g'(0) \leq 0, \quad g'(\pi/3) \ge 0,
\eeq
and
\beq
\lb{lultimasega}
g'_-(\theta_i) > g'_+(\theta_i), \quad \forall \theta_i.
\eeq

\end{theorem}

\begin{figure}[!htb]
\begin{center}
\vspace*{3mm}
\includegraphics[width=7cm]{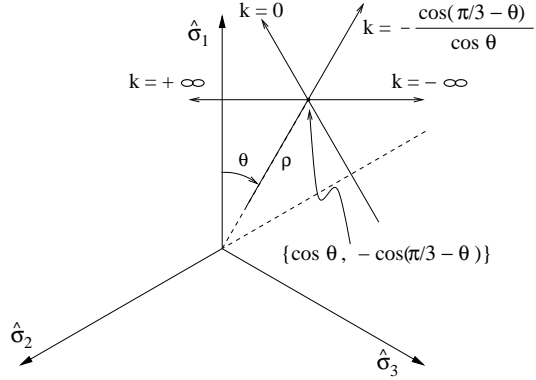}
\caption{\footnotesize 
Bundle of lines $\{ \cos\theta + \epsilon,-\cos(\pi/3-\theta) + k \epsilon \}$ in the deviatoric plane. Note that 
$k$ represents the slope of the lines in the nonorthogonal reference system $\hat{\sigma}_1, \hat{\sigma}_2$.}
\label{figD}
\end{center}
\end{figure}

\begin{proof}
Conditions (\ref{curvaturapoi}) have been already proven in Theorem  \ref{cornerone} and do not need further explanation. We therefore 
restrict our attention to the singular points $\theta_i$, to derive condition (\ref{lultimasega}). 

We consider a generic point in the first $\pi/3$-sector of the deviatoric plane (taken clockwise from axis $\hat{\sigma}_1$; the proof can be 
easily extended to the other sectors), $2q/3\{\cos\theta, -\cos(\pi/3-\theta)\}$, and the parametric representation (with the 
parameter $\epsilon$) of all lines of slope $k$ through this point (Fig.\ \ref{figD})
\beq
\lb{lineaaaaa}
\left\{
\frac{2}{3} q \cos\theta +\epsilon, -\frac{2}{3} q \cos\left(\frac{\pi}{3}-\theta \right) + k \epsilon
\right\}.
\eeq
The derivative of the restriction $h$ of $q/g(\theta)$ to this line is again given by eqn.\ (\ref{derivatazza}), with all the functions calculated 
at points (\ref{lineaaaaa}). Taking the limit values at $\epsilon = 0$,
\beq
\nabla q = \left\{ \frac{3}{2} \cos\theta-\frac{\sqrt{3}}{2}\sin\theta, -\sqrt{3} \sin\theta \right\}, \quad
\nabla \theta = \frac{\sqrt{3}}{q}\hat{H} \left\{
\cos\left( \frac{\pi}{3}-\theta\right), \cos\theta \right\},
\eeq
and noting that $\hat{H} = -1$ in the $\pi/3$-sector under consideration (see Fig.\ \ref{figA}), we obtain
\beq
\lb{machecazzoe}
\lim_{\epsilon\rightarrow 0^\pm} h'(\epsilon) = \frac{\sqrt{3}}{2\,g(\theta)} \left[\sqrt{3} \cos\theta-(1+2k)\sin\theta \right]
- \frac{\sqrt{3}}{2}\frac{g'_{(*)}(\theta)}{g^2(\theta)}\hat{H} \left[(1+2k)\cos\theta +\sqrt{3}\sin\theta \right],
\eeq	
where 
\beq
(*) = \left\{
\barr{lll}
\pm & \mbox{for}  & \ds -\infty < k < -\frac{\cos\left(\pi/3-\theta\right)}{\cos\theta}, \\[5mm]
\mp & \mbox{for}  & \ds -\frac{\cos\left(\pi/3-\theta\right)}{\cos\theta} < k < +\infty. 
\earr
\right.
\eeq
Using eqn.\ (\ref{machecazzoe}) into condition $h'_-(0) < h'_+(0)$ yields in both cases inequality (\ref{lultimasega}).

\end{proof}

%%%%%%%%%%%%%%%%%%%%%%%%%%%%%%%%%%%%%%%%%%%%%%%%%%%%%%%%%%%%%%%%%%%%%%
%%%%%%%%%%%%%%%%%%%%%%%%%%%%%%%%%%%%%%%%%%%%%%%%%%%%%%%%%%%%%%%%%%%%%%

We are now in a position to state the generalization of the Proposition 1 given by Bigoni and Piccolroaz (2004) to yield surfaces with 
corners.

\begin{theorem}
\lb{generalissimofranco} 
\underline{Convexity of piecewise-smooth yield function vs. convexity of the yield  surface.}

Convexity of the yield {\it function} (\ref{funzionazza}) is equivalent to convexity of the me\-ri\-dian and deviatoric sections of the 
corresponding yield {\it surface} in the Haigh-Westergaard representation. In symbols:
\beq
{\rm convexity~of~} F(\bsigma) = f(p) + \frac{q}{g(\theta)} \Longleftrightarrow 
\left\{
\barr{ll}
f''\geq0, \\[5mm]
g^2 + 2g'^2 - gg'' \geq 0, \quad \forall \theta \in (0,\pi/3)-\{\theta_i\}  \\[5mm]
g'(0) \leq 0, \quad g'(\pi/3) \ge 0, \\[5mm]
g'_-(\theta_i) > g'_+(\theta_i), \quad \forall \theta_i.
\earr
\right.
\eeq
where $g(\theta)$ is a continuous and strictly positive function in $[0,\pi/3]$, twice-differentiable almost everywhere in $(0,\pi/3)$, and 
$\theta_i$ denotes the singular points of $g(\theta)$.
\end{theorem}

\begin{proof}
The proof follows directly from Lemma \ref{lemmino} and Theorems \ref{hillhill} and \ref{supercorner}.
\end{proof}

\section{Conclusions}

Yield surfaces used in elastoplasticity theory often have corners. For these nonsmooth functions, we have given in this paper a general 
theorem providing necessary and sufficient conditions for the equivalence between the convexity of the deviatoric yield function and its 
representation as a surface in the Haigh-Westergaard stress space. This theorem is useful for the definition of new yield function or 
transformation function for phase-transforming materials. We have also provided a generalization to nonsmoothness of a theorem relating 
convexity of a scalar isotropic function of tensorial variable to the convexity of the corresponding functions of the tensor principal values. 
This can find applications in the formulation of nonsmooth-convex elastic potential energy functions.

\section*{\normalsize{Acknowledgements}}

DB acknowledges financial support of PRIN grant n. 2007YZ3B24
"Multi-scale Problems with Complex Interactions in Structural Engineering" financed by Italian Ministry of University
and Research.

\clearpage
%%%%%%%%%%%%%%%%%%%%%%%%%%%%%%%%%%%%%%%%%%%%%%%%%%%%%%%%%%%%%%%%%%%%%%%
%Appendix
\appendix
\renewcommand{\theequation}{\thesection.\arabic{equation}}

\section{APPENDIX}
\setcounter{equation}{0}

\subsection{Proof of the auxiliary property of the scalar product of two symmetric tensors}
\lb{theproof}

We provide the proof of the auxiliary property of the scalar product of two symmetric tensors, which is often used
(among others, by Ogden, 1984). 
The property has been noticed by Hill (1968), 
who did not provide a complete proof (which is only sketched in a footnote), perhaps because of a lack of space.
We were not able to find a proof of the property anywhere.

\begin{theorem}
Let $\bA,\bB$ be two symmetric tensors. Then, denoting by $\alpha_1,\alpha_2,\alpha_3$ and $\beta_1,\beta_2,\beta_3$ the eigenvalues of $\bA$ 
and $\bB$, respectively,
\beq
\bA \cdot \bB \leq \alpha_1\beta_1 + \alpha_2\beta_2 + \alpha_3\beta_3,
\eeq
given that the eigenvalues of the two tensors are numbered in the same algebraic order.
\end{theorem}

\begin{proof}
Given the eigenvalues of the two tensors, we keep the eigenvectors $\ba_1,\ba_2,\ba_3$ of $\bA$ fixed and seek for the maximum of $\bA \cdot \bB$ 
as the eigenvectors $\bb_1,\bb_2,\bb_3$ of $\bB$ rotate with respect to $\ba_1,\ba_2,\ba_3$. Therefore, the problem can be formulated in terms of 
the following optimization problem 
\beq
\max_{\bb_1,\bb_2,\bb_3} \bA \cdot \bB,
\eeq
with the constraint that $(\bb_1,\bb_2,\bb_3)$ be an orthonormal basis, 
\beq
\lb{const}
\bb_M \cdot \bb_N = \delta_{MN}, 
\eeq
where $\delta_{MN}$ is the Kronecker symbol.

This optimization problem can be solved using Lagrangean multipliers, so that we maximize the function
\beq
\barr{ll}
\bA \cdot \bB = & \alpha_1\beta_1 (\ba_1 \cdot \bb_1)^2 + \alpha_1\beta_2 (\ba_1 \cdot \bb_2)^2 + \alpha_1\beta_3 (\ba_1 \cdot \bb_3)^2 \\[3mm]
& + \alpha_2\beta_1 (\ba_2 \cdot \bb_1)^2 + \alpha_2\beta_2 (\ba_2 \cdot \bb_2)^2 + \alpha_2\beta_3 (\ba_2 \cdot \bb_3)^2 \\[3mm]
& + \alpha_3\beta_1 (\ba_3 \cdot \bb_1)^2 + \alpha_3\beta_2 (\ba_3 \cdot \bb_2)^2 + \alpha_3\beta_3 (\ba_3 \cdot \bb_3)^2 \\[3mm]
& + \Lambda_1 (\bb_1 \cdot \bb_1 - 1) + \Lambda_2 (\bb_2 \cdot \bb_2 - 1) + \Lambda_3 (\bb_3 \cdot \bb_3 - 1) \\[3mm]
& + \Lambda_4 (\bb_1 \cdot \bb_2) + \Lambda_5 (\bb_2 \cdot \bb_3) + \Lambda_6 (\bb_3 \cdot \bb_1), 
\earr
\eeq
as a function of $\bb_1,\bb_2,\bb_3$ and the Lagrangean multipliers $\Lambda_i$ ($i=1,...,6$), thus obtaining
\beq
\lb{syst}
\barr{l}
\ds \deriv{\bA \cdot \bB}{\bb_1} = 2 \beta_1 \bA \bb_1 + 2 \Lambda_1 \bb_1 + \Lambda_4 \bb_2 + \Lambda_6 \bb_3 = 0, \\[3mm]
\ds \deriv{\bA \cdot \bB}{\bb_2} = 2 \beta_2 \bA \bb_2 + 2 \Lambda_2 \bb_2 + \Lambda_4 \bb_1 + \Lambda_5 \bb_3 = 0, \\[3mm]
\ds \deriv{\bA \cdot \bB}{\bb_3} = 2 \beta_3 \bA \bb_3 + 2 \Lambda_3 \bb_3 + \Lambda_5 \bb_2 + \Lambda_6 \bb_1 = 0, 
\earr
\eeq
together with the constraints \eq{const}. 

In the case of distinct eigenvalues $\beta_1,\beta_2,\beta_3$, the system \eq{syst} is satisified if and only if
\beq
\bb_M \cdot \bA \bb_N = 0, \quad \text{for} \quad M \neq N,
\eeq
and thus if and only if $\bb_1,\bb_2,\bb_3$ are eigenvectors of $\bA$.
This proves that the extreme values of $\bA \cdot \bB$ are attained when the two tensors are coaxial. The maximum is then selected from six 
possibilities.

In the case $\beta_1 = \beta_2 \neq \beta_3$, the same line of thought used above allows us to conclude that the extreme values of $\bA \cdot \bB$ 
are attained when $\bb_3$ is an eigenvector of $\bA$, in which case the two tensors $\bA$ and $\bB$ are coaxial and, choosing $\bb_3 \equiv \ba_3$, 
$\bA \cdot \bB = (\alpha_1 + \alpha_2)\beta_1 + \alpha_3\beta_3$.

The case $\beta_1 = \beta_2 = \beta_3$ is trivial. The two tensors $\bA$ and $\bB$ are coaxial and the scalar product is 
$\bA \cdot \bB = (\alpha_1 + \alpha_2 + \alpha_3)\beta_1$.
\end{proof}

\subsection{The convexity condition given by Laydi and Lexcellent (2009)}
\lb{lexcellent}

\begin{figure}[!htb]
\begin{center}
\vspace*{3mm}
(a) \includegraphics[width=7.3cm]{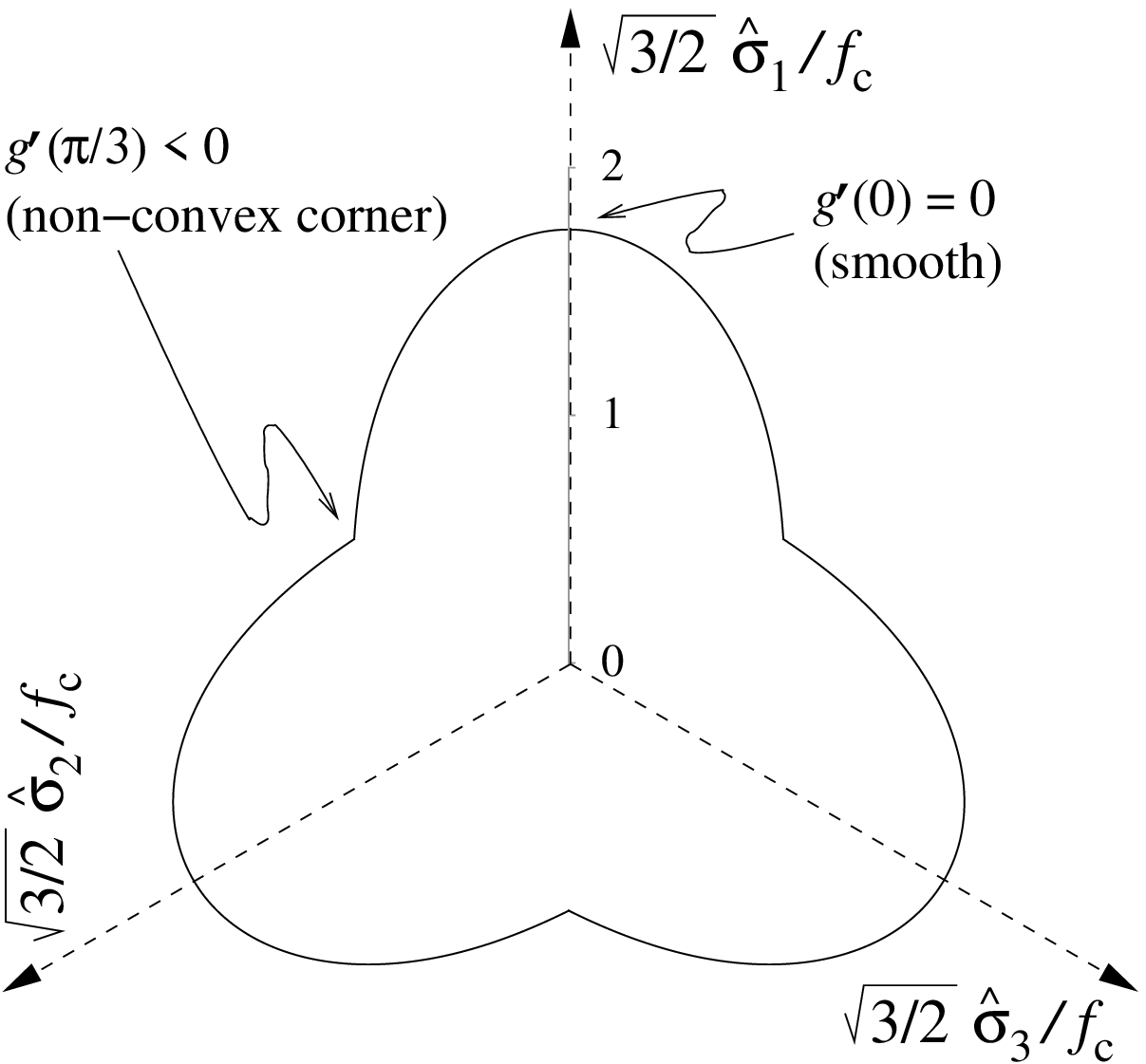}
\hspace{0mm}
(b) \includegraphics[width=6cm]{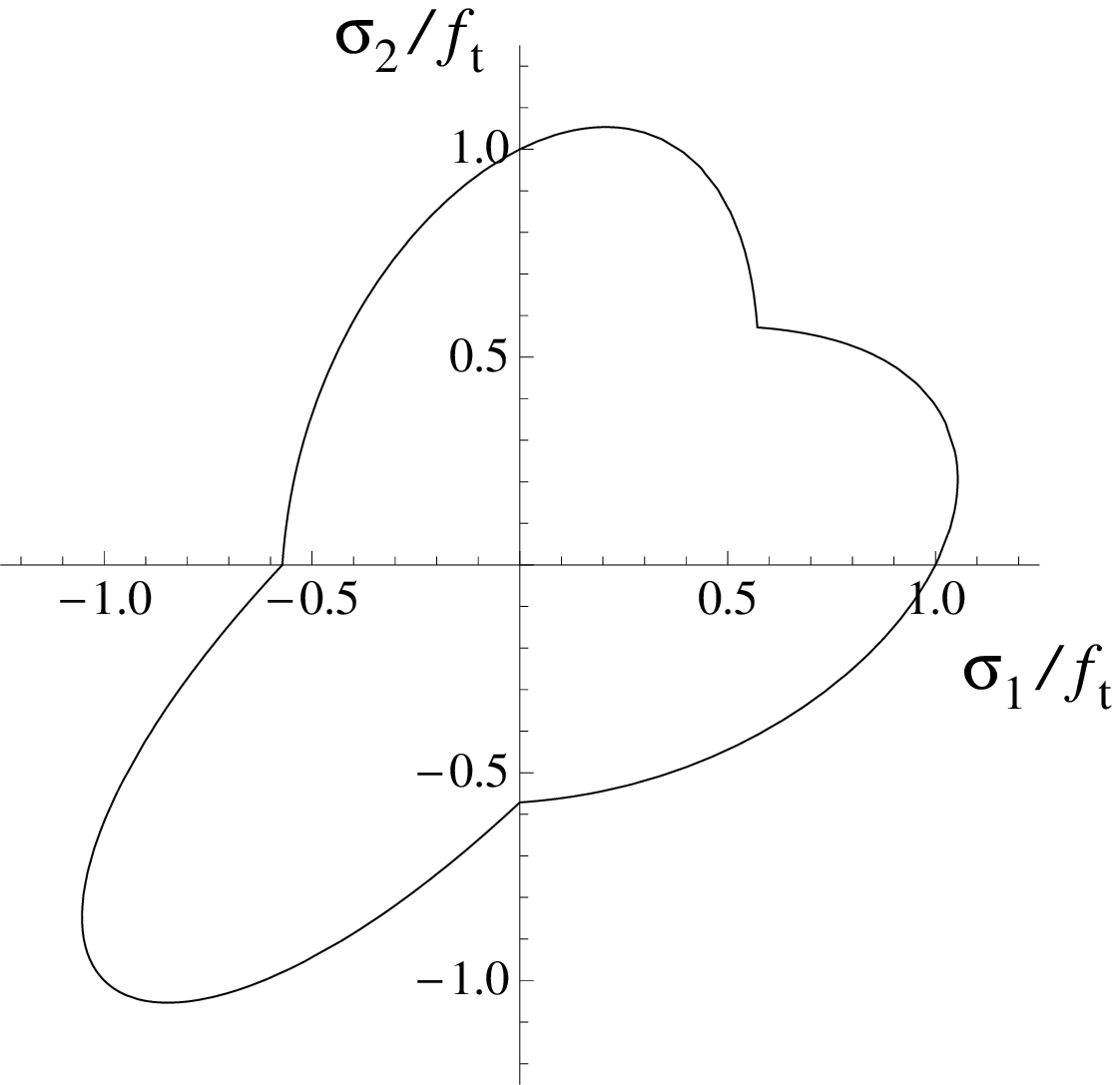}
\caption{\footnotesize The example by Laydi and Lexcellent (2009) showing the possibility of describing corners 
within the class of functions (\ref{funzionazza}) with the choice (\ref{acazzi}). Since $g'(\pi/3) < 0$, hypotheses of 
Theorem \ref{cornerone} are violated and indeed 
the deviatoric section
of the yield surface has reentrant corners at $\theta=\pi/3$.
(a) Deviatoric section. (b) Yield
surface in the biaxial plane $\sigma_1/f_t$
vs. $\sigma_2/f_t$, with $\sigma_3 = 0$.}
\label{fig02a}
\end{center}
\end{figure}

Laydi and Lexcellent (2009) have shown, with the example reported in Fig.\ \ref{fig02a}, that the convexity conditions given by 
Bigoni and Piccolroaz (2004) does not cover yield surfaces with corners. In fact, by selecting within the class (\ref{funzionazza}) 
the following function
\beq
\lb{acazzi}
F(\bsigma) = -f_t + q/g(\theta), \quad
g^{-1}(\theta) = 2 - \cos^2 \theta,
\eeq
the Bigoni and Piccolroaz (2004) conditions are satisfied, but, although $g'(0) = 0$, the yield surface has concave corners at $\theta = \pi/3$, 
$g'(\pi/3) < 0$ [instead of $g'(\pi/3) > 0$, corresponding to convex corners], see Fig.\ \ref{fig02a}.

%%%%%%%%%%%%%%%%%%%%%%%%%%%%%%%%%%%%%%%%%%%%%%%%%%%%%%%%%%%%%%%%%%%%%%
%%%%%%%%%%%%%%%%%%%%%%%%%%%%%%%%%%%%%%%%%%%%%%%%%%%%%%%%%%%%%%%%%%%%%%

Laydi and Lexcellent (2009) incorrectly argued that the Bigoni and Piccolroaz (2004) proposition on convexity had flaws, while the
problem lies only in the fact that the deviatoric section of the yield surface described by eqns.\ (\ref{acazzi}) has corners, a case 
which is not covered by the Bigoni and Piccolroaz (2004) proposition and has been addressed in the present paper.

Laydi and Lexcellent (2009) also provided sufficient conditions for convexity of the deviatoric section of a smooth yield surface. These conditions in our notation 
read
\beq
\lb{cazzoni2}
\left\{
\barr{l}
\ds -\cos\theta (gg') + \sin\theta g^2 \geq 0, \\[5mm]
\ds \cos\theta (gg') + \sin\theta (2g'^2 - gg'') \geq 0,
\earr
\right.
\eeq
for all $\theta \in [0,\pi/3]$.

However, these conditions are neither necessary,  nor sufficient for deviatoric sections with corners, while the correct, 
necessary and sufficient conditions are those specified by Theorem \ref{cornerone}. 
To fully justify this statement, we provide the two counter-examples below.

\underline{Counter-example 1}: Conditions \eq{cazzoni2} are not necessary for convexity of yield functions, even 
with smooth deviatoric section. 

This is made clear by the following counter-example (taken from eqn.\ \eq{effedip}$_2$ with $\beta = 0.5$ and $\gamma = 0.99$):
\beq
\lb{count1}
F(\bsigma) = -\frac{f_c}{g(\pi/3)} + \frac{q}{g(\theta)}, \quad
\frac{1}{g(\theta)} = \cos{\left[ 0.5\, \frac{\pi}{6} - \frac{\cos^{-1}\left(0.99 \cos{3 \theta}\right)}{3}\right]}. 
\eeq

\begin{figure}[!htb]
\begin{center}
\vspace*{3mm}
(a) \includegraphics[width=7.3cm]{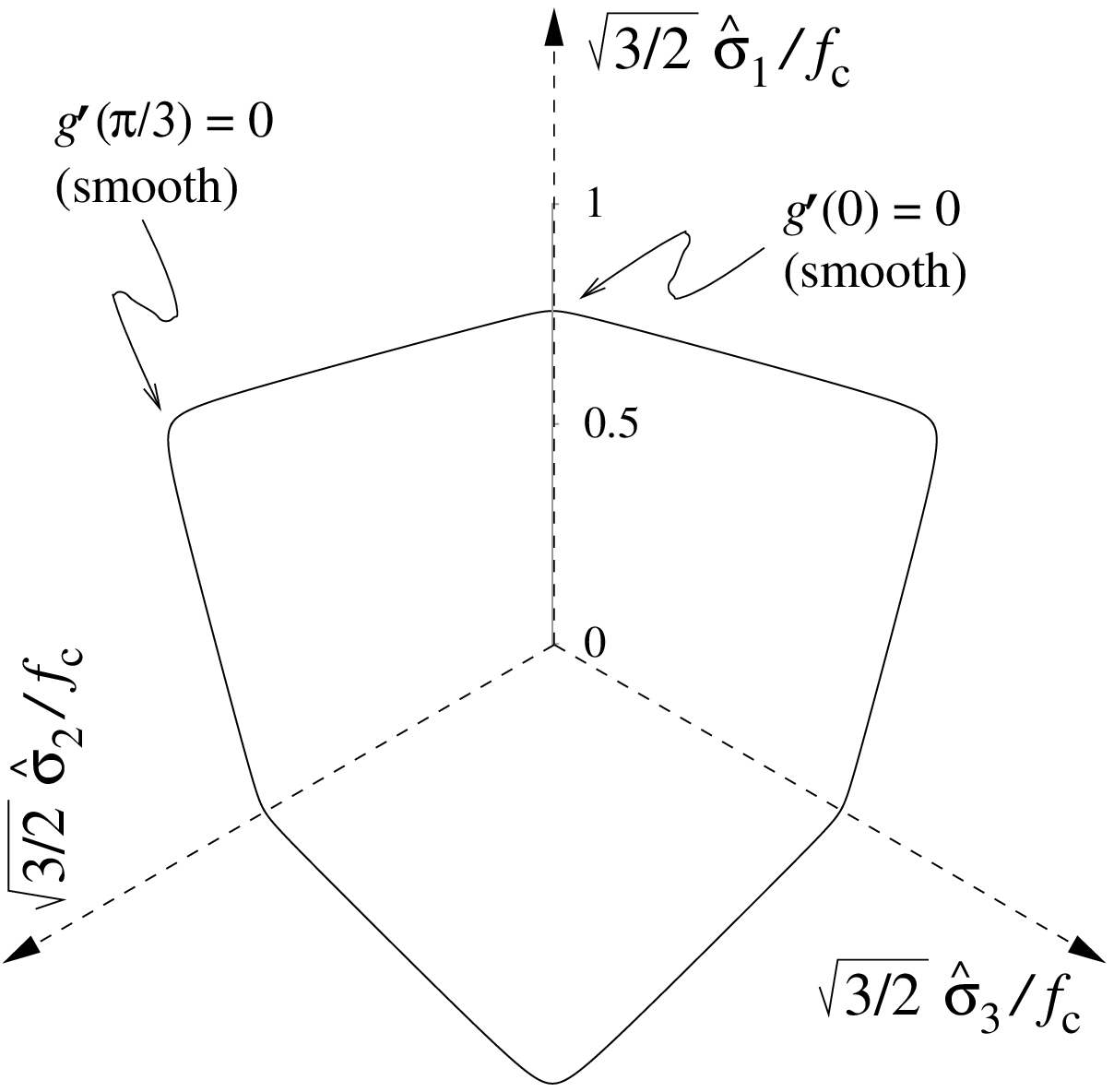}
\hspace{0mm}
(b) \includegraphics[width=6cm]{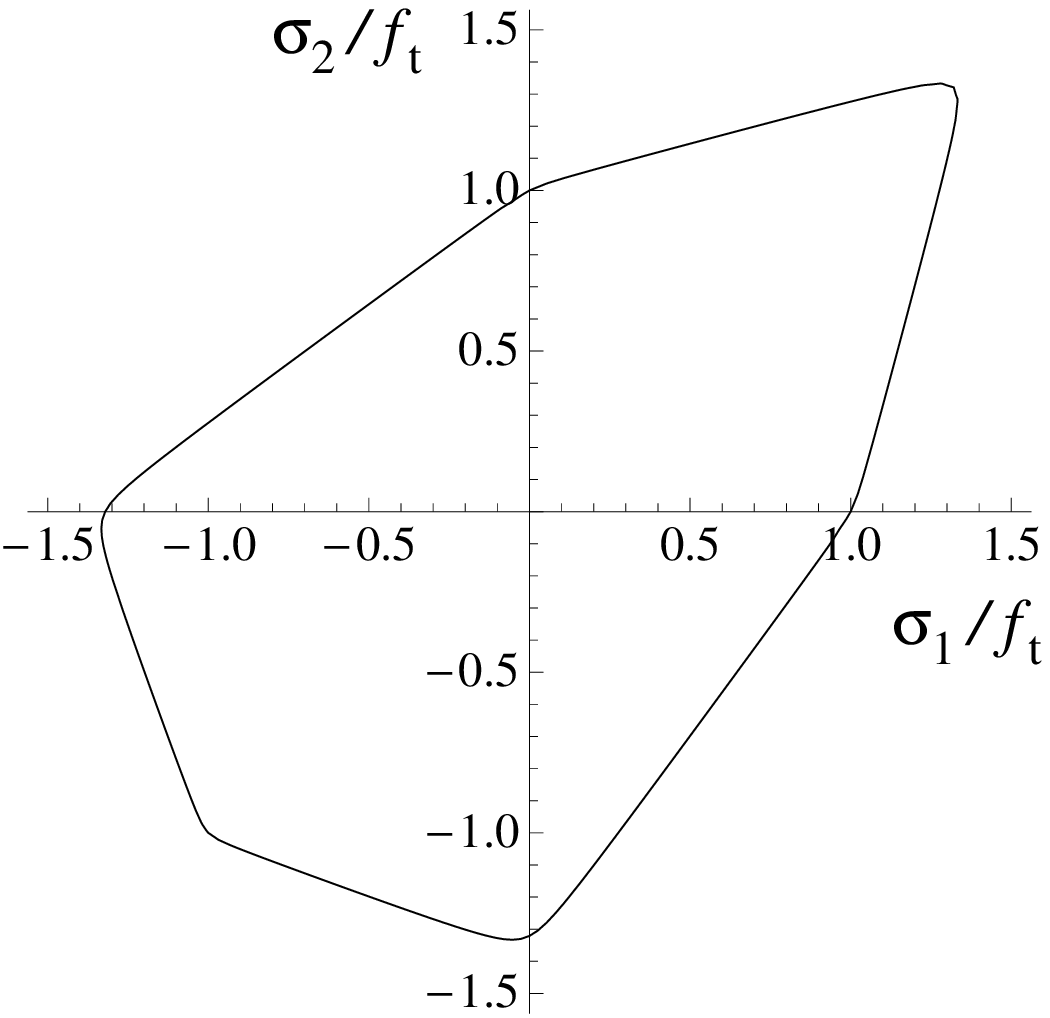}
\caption{\footnotesize An example of convex and smooth yield surface, corresponding to eqns.\ \eq{count1}, not satisfying the Laydi and Lexcellent (2009) conditions \eq{cazzoni2}.
(a) Deviatoric section. (b) Yield surface in the biaxial plane $\sigma_1/f_t$ vs. $\sigma_2/f_t$, with $\sigma_3 = 0$.}
\label{fig03a}
\end{center}
\end{figure}

The deviatoric shape function \eq{count1}$_2$ corresponds to a smooth and convex deviatoric section, see Fig.\ \ref{fig03a}, but it is easy to show 
that it {\em does not} satisfy the condition \eq{cazzoni2}$_2$.

%%%%%%%%%%%%%%%%%%%%%%%%%%%%%%%%%%%%%%%%%%%%%%%%%%%%%%%%%%%%%%%%%%%%%%
%%%%%%%%%%%%%%%%%%%%%%%%%%%%%%%%%%%%%%%%%%%%%%%%%%%%%%%%%%%%%%%%%%%%%%

\underline{Counter-example 2}: Conditions \eq{cazzoni2} are not sufficient for convexity of deviatoric sections 
with corners. 

This is made clear by the following counter-example:
\beq
\lb{count2}
F(\bsigma) = -\frac{f_c}{g(\pi/3)} + \frac{q}{g(\theta)}, \quad
g(\theta) = \theta^2 - 0.8\, \theta^4 - \theta \sin\theta + 1. 
\eeq

\begin{figure}[!htb]
\begin{center}
\vspace*{3mm}
(a) \includegraphics[width=7.3cm]{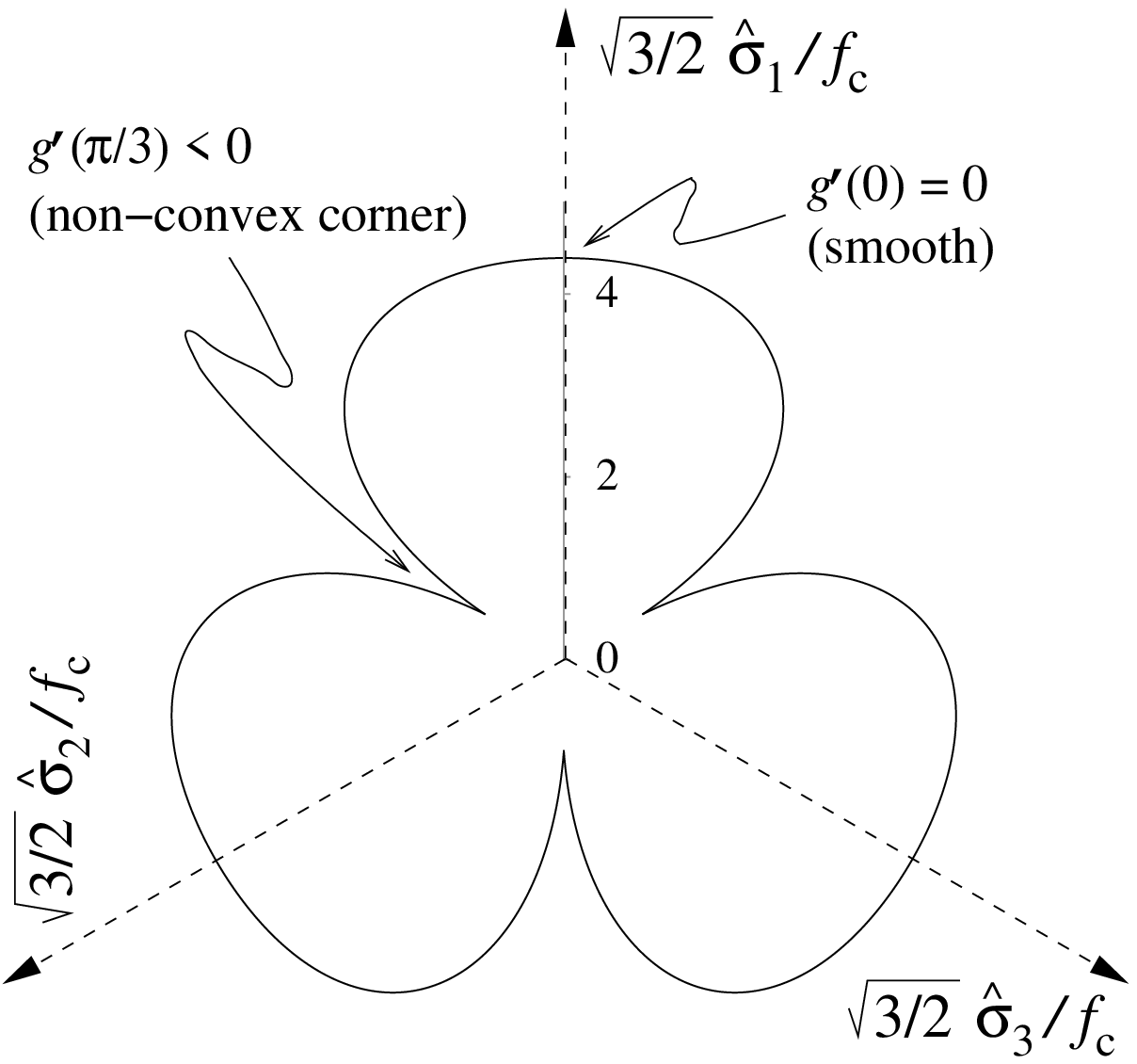}
\hspace{0mm}
(b) \includegraphics[width=6cm]{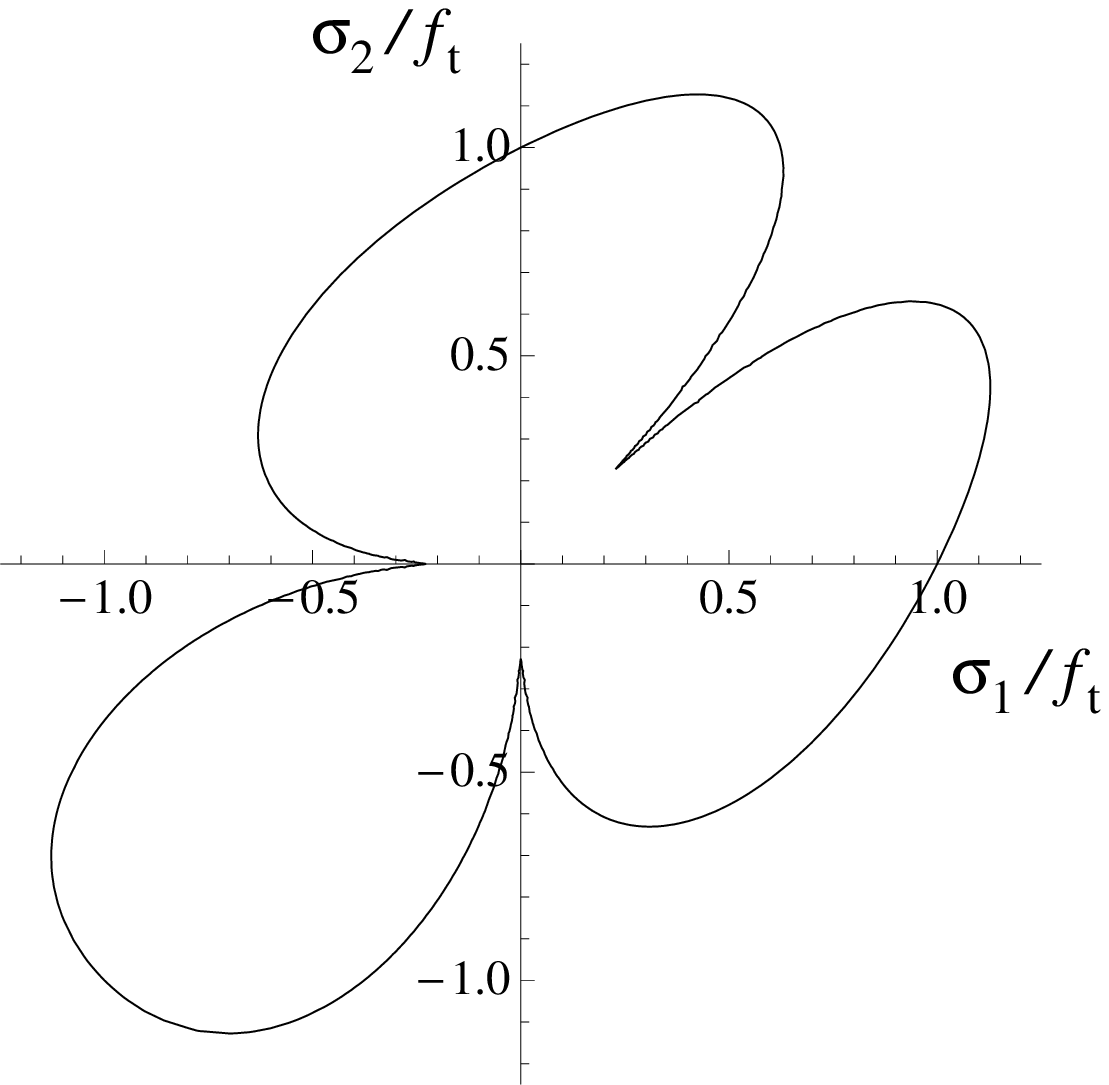}
\caption{\footnotesize An example of non-convex yield surface, corresponding to eqns.\ \eq{count2}, satisfying the Laydi and Lexcellent (2009) conditions \eq{cazzoni2}.
(a) Deviatoric section. (b) Yield surface in the biaxial plane $\sigma_1/f_t$ vs. $\sigma_2/f_t$, with $\sigma_3 = 0$.}
\label{fig04a}
\end{center}
\end{figure}

The deviatoric shape function \eq{count2}$_2$ corresponds to a non-convex deviatoric section, see Fig.\ \ref{fig04a}, but it is easy to show 
that it {\em does} satisfy both conditions \eq{cazzoni2}.

%%%%%%%%%%%%%%%%%%%%%%%%%%%%%%%%%%%%%%%%%%%%%%%%%%%%%%%%%%%%%%%%%%%%%%
%%%%%%%%%%%%%%%%%%%%%%%%%%%%%%%%%%%%%%%%%%%%%%%%%%%%%%%%%%%%%%%%%%%%%%

\end{document}